\documentclass[%
 reprint,
 amsmath,amssymb,
 aps,
]{revtex4-2}
\usepackage{xcolor}
\usepackage{float}
\usepackage{amsmath}
\usepackage{multirow}
\usepackage{subcaption}
\usepackage{graphicx}
\usepackage{dcolumn}
\usepackage{bm}

\begin{document}

\title{Density and Affinity Dependent Social Segregation and Arbitrage Equilibrium in a Multi-class Schelling Game}

\author{Venkat Venkatasubramanian}%
 \email{venkat@columbia.edu}
\affiliation{%
Complex Resilient Intelligent Systems Laboratory, Department of Chemical Engineering, Columbia University, New York, NY 10027, U.S.A.
}%

\author{Jessica Shi}%
\affiliation{%
Department of Industrial Engineering and Operations Research, Columbia University, New York, NY 10027, U.S.A.}%

\author{Leo Goldman}%
\affiliation{%
Department of Computer Science and Engineering, Columbia University, New York, NY 10027, U.S.A.}%

\author{Arun Sankar E.M.}%
\affiliation{%
Complex Resilient Intelligent Systems Laboratory, Department of Chemical Engineering, Columbia University, New York, NY 10027, U.S.A.}

\author{Abhishek Sivaram}
\affiliation{Complex Resilient Intelligent Systems Laboratory, Department of Chemical Engineering, Columbia University, New York, NY 10027, U.S.A.}

\date{\today}

\begin{abstract}
Contrary to the widely believed hypothesis that larger, denser cities promote socioeconomic mixing, a recent study~\cite{nilforoshan2023human} reports the opposite behavior, i.e. more segregation. Here, we present a game-theoretic model that predicts such a density-dependent segregation outcome in both one- and two-class systems. The model provides key insights into the analytical conditions that lead to such behavior. Furthermore, the arbitrage equilibrium outcome implies the equality of effective utilities among all agents. This could be interpreted as all agents being equally "happy" in their respective environments in our ideal society. We believe that our model contributes towards a deeper mathematical understanding of social dynamics and behavior, which is important as we strive to develop more harmonious societies. 
\end{abstract}
\maketitle

\section{Introduction}

The increasing economic inequality and the related increase in socioeconomic segregation are of concern in many societies. In the U.S., for example, this segregation leads to wide variations in lifestyles and life outcomes. This lack of mixing and social interactions across the socioeconomic spectrum contributes to social tensions and political polarization. It is generally believed that in large, dense, urbanized environments, the diverse happenstance of social interactions among people leads to more mixing and less segregation. However, in a recent study, Nilforoshan et al.\cite{nilforoshan2023human} report that large cities increase rather than reduce socioeconomic segregation, as they can offer a greater choice of differentiated spaces aimed at specific socioeconomic groups. They observe that "The consistent result that larger, denser cities are more segregated runs counter to the hypothesis that such cities promote socioeconomic mixing by attracting diverse individuals and constraining space in ways that oblige them to encounter one
other. Our results support the opposite hypothesis: big cities allow their inhabitants to seek out people who are more like themselves." \\

In this work, we present a mathematical model and its agent-based simulation that demonstrate this density-dependent segregation outcome. Our work is related to Schelling's seminal game-theoretic model of social segregation\cite{schelling}. It is an agent-based model in which agents move according to a specified utility function that depends on their neighbors. He showed that although agents may only have a mild preference, they still choose to segregate into neighborhoods of similar individuals over time. Schelling's model was originally proposed to study social segregation phenomena but can also be related to phase separation in chemistry and physics~\cite{vinkovic2006physical, gauvin2, dall2008statistical, avetisov2018phase, grauwin}. Hence, in this paper, we use the terms phase separation and social segregation interchangeably. \\

It turns out that this connection between two apparently very different domains, sociology/economics and physics/chemistry, is not just coincidental. There exists a deep connection between game theory (the mathematical framework for modeling phenomena in sociology/economics) and statistical mechanics (the mathematical framework for modeling equilibrium phenomena in physics/chemistry) that was identified by Venkatasubramanian \cite{venkat2017book}. Inspired by this insight, he developed a novel analytical framework, called \textit{statistical teleodynamics}, which is a synthesis of the central concepts and techniques of statistical mechanics and population game theory \cite{venkat2015howmuch, venkat2017book, venkat2022unified}. In this paper, we use this framework to model the dynamics of socioeconomic segregation, which reveals interesting insights. We study Schelling-like systems with one-class and two-class models to determine the analytical conditions under which they would undergo socioeconomic segregation. We extend the analysis first introduced by Venkatasubramanian et al.\cite{venkat2022unified} to two-class systems and a wider range of parameters. 

\section{Mathematical Formulation: Pursuit of Maximum Utility and Arbitrage Equilibrium}

Our goal is to understand the fundamental principles and mechanisms of self-organization of goal-driven social agents. Toward that end, we develop simple models that offer an appropriate coarse-grained description of the system. Unlike atoms and molecules, social agents do not behave precisely and predictably. Therefore, we have deliberately tried to keep the models as simple as possible and not be restricted by system-specific details and nuances without losing key insights and relevance to empirical phenomena~\cite{venkat2022unified}.\\

We also wish to stress that the spirit of our modeling is similar to that of the ideal gas or the Ising model in statistical thermodynamics. Just as real molecules are not point-like objects or are devoid of intermolecular interactions, as assumed in the ideal gas model in statistical mechanics, we make similar simplifying assumptions about the social agents in our model. These can be relaxed to make them more realistic in subsequent refinements like, for example, van der Walls did in thermodynamics. Ideal versions serve as useful starting points and reference states for developing more comprehensive models of self-organization in sociological systems.\\

The central theme in our theory is that goal-driven social agents constantly pursue maximum utility by jockeying for better positions via self-organization. So, we formulate the problem by first defining the \emph{effective utility}, $h_{i}$, for an agent in state $i$. The effective utility is the net sum of the \textit{benefits} minus the \textit{costs} for the agents. Social agents constantly make benefit-cost trade-offs in their self-driven dynamical behavior to improve their \textit{socioeconomic fitness}, i.e. the effective utility. This results in a delicate dynamic balance of the benefits of aggregation versus the costs of overcrowding the agents. In other words, the benefits of \textit{cooperation} are balanced with the costs of \textit{competition}. \\

Furthermore, driven by natural instincts, agents also balance two competing strategies - \textit{exploitation} and \textit{exploration}. Exploitation takes advantage of the opportunities in the immediate, local neighborhood. On the other hand, exploration examines possibilities outside.\\

We believe that this combination of two main strategies, namely, the benefit-cost trade-offs of the \textit{cooperation-competition} strategy with an \textit{exploitation-exploration} strategy, is a fundamental and universal evolutionary mechanism found in most living systems.\\

We motivate our model by initially considering a one-class system, which is simpler to start with. Here, all agents belong to the same socioeconomic category. As is generally done in Schelling games, we also model the space where the agents operate as a large lattice $L$ of local neighborhoods or blocks, each with $M$ sites that agents can occupy. There are $Q$ such blocks, $QM$ sites, and a total of $N$ agents, with an average agent density of $\rho_0 = N/(QM)$. The state of an agent is defined by specifying the block $i$ in which it is located, and the state of the system is defined by specifying the number of agents, $N_i$, in block $i$, for all blocks ($i \in \{1, \dots, Q\}$). The density of the agents block $i$ is given by $\rho_i = N_i/N$. Let block $i$ also have $V_i$ vacant sites, so $V_i = M - N_i$. This approach is an extension of our recent model developed for a Schelling-like game scenario \cite{venkat2022unified}. \\

We further formulate the problem by defining the effective utility, $h_i$, for the one-class agents in block $i$, which agents try to maximize by moving to better locations (i.e., other blocks), if possible. The effective utility is the net sum of the benefits minus the costs and has four components. The first is that an agent prefers to have more agents in its neighborhood, as this aggregation improves its socioeconomic quality of life.  Therefore, this \textit{affinity benefit} term, representing cooperation among agents, is proportional to the number of agents in its neighborhood. We model this as $\alpha N_i$, where $\alpha >0$ is a parameter. \\

\begin{figure}[ht]
\includegraphics[width=0.5\textwidth]{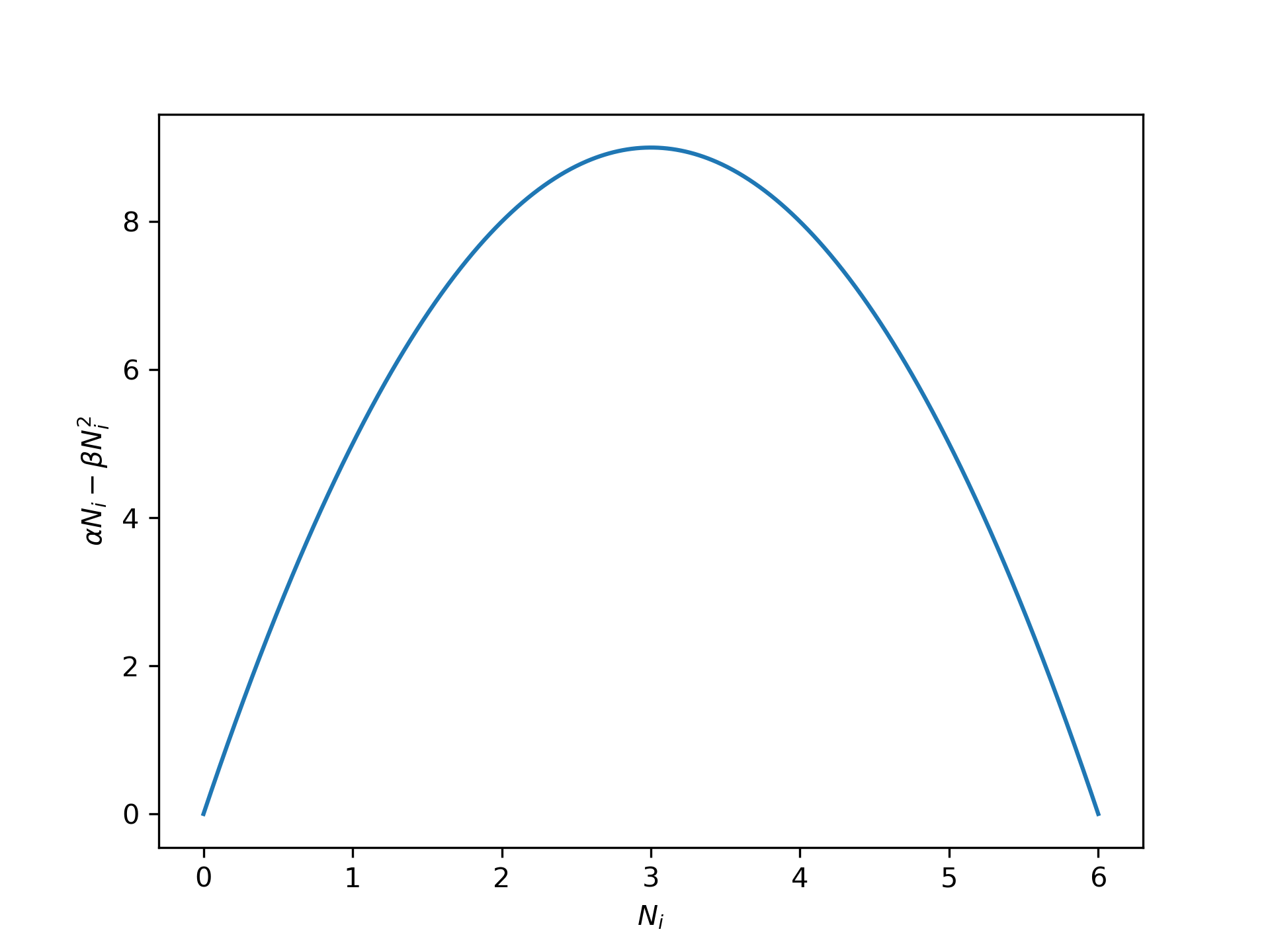}
\caption{Net benefit of a resource for $\alpha N_i - \beta {N_i}^2$ ( $\alpha=6$ , $\beta =1$)}
\label{fig:U_curve}
\end{figure}

However, this affinity benefit comes with a cost.  As more and more agents aggregate, this overcrowding results in a \textit{congestion cost} term. As Venkatasubramanian explains \cite{venkat2017book}, the resulting net benefit (= benefit - cost) function has an inverted U-like profile (see Fig.~\ref{fig:U_curve}). This profile is found in many net benefit vs. resource relationships in the real world. As one consumes a resource, it initially leads to increasing net benefit; but after a point, the cost of the resource goes up more quickly than the benefit, thus resulting in decreasing net benefit. The simplest model of this is a quadratic function, $\alpha N_i - \beta {N_i}^2$, with the quadratic term $-\beta N_i^2$ ($\beta >0$) modeling the congestion cost.  \\

Regarding exploration, agents derive a benefit by having a large number of vacant sites to potentially move to in the future should such a need arise. This is the instinct to explore other opportunities, as new vacant sites are potentially new sources of socioeconomic benefits. We call this the \textit{option benefit} term, as agents have the option to move elsewhere if needed. Again, following Venkatasubramanian~\cite{venkat2017book, kanbur2020occupational, venkat2022unified}, we model this as $\gamma \ln (M-N_i)$,  $\gamma >0$. The logarithmic function captures the diminishing utility of this option, a commonly used feature in economics and game theory. As before, this benefit is also associated with a cost as a result of competition among agents for these vacant sites. We model this \textit{disutility of competition} as $-\delta \ln N_i$, $\delta >0$ \cite{venkat2015howmuch, venkat2017book, kanbur2020occupational}.\\

Combining these four components, we have the following effective utility function $h_i$ for the agents in block $i$ as, 
\begin{eqnarray}
    h_i(N_i) = \alpha N_i  - \beta N_i^2 + \gamma \ln(M - N_i) - \delta \ln N_i
\end{eqnarray}

Intuitively, the first two terms in the equation model the benefit-cost trade-off in the exploitation behavior, while the last two model a similar trade-off in exploration. \\

We can set $\delta = 1$ without any loss of generality. In addition, we set $\gamma = 1$ to gain analytical simplicity, but this can be relaxed later if necessary. So we now have

\begin{eqnarray}
    h_i(N_i) = \alpha N_i  - \beta N_i^2 + \ln(M - N_i) - \ln N_i
    \label{eq:Utility_N_i}
\end{eqnarray}

Rewriting this in terms of the density ($\rho_i$) of agents in block $i$, $\rho_i = N_i/M$, and absorbing the constant $M$ into $\alpha$ and $\beta$, we have

\begin{eqnarray}
    h_i(\boldsymbol \rho) = \alpha \rho_i - \beta \rho_i^2 + \ln (1- \rho_i) - \ln \rho_i
    \label{eq:Utility}
\end{eqnarray}

For simplicity, we define $u(\rho_i) = \alpha \rho_i - \beta \rho_i^2$. Therefore, the potential $\phi(\boldsymbol{\rho})$ becomes 

\begin{eqnarray}
    \phi(\boldsymbol{\rho}) &=& \sum_{i=1}^n \int h_i(\boldsymbol{x}) d x_i = \frac{M}{N}\sum_{i=1}^n \int h_i(\boldsymbol{\rho}) d \rho_i \nonumber\\
    &=&\frac{M}{N}\sum_{i=1}^n \int_0^{\rho_i}\left[ u(\rho) + \ln (1-\rho) - \ln \rho \right] d \rho \nonumber\\
    \label{eq:schelling-potential}
\end{eqnarray}
One can generalize the discrete formulation to a continuous one by replacing $\rho_i$ by $\rho(r)$, where the density is a continuous function of radius $r$ of the neighborhood as demonstrated by Sivaram and Venkatasubramanian~\cite{venkat2022garuds} in the self-organized flocking behavior of birds. \\

Now, according to the theory of potential games~\cite{sandholm2010population}, an \textit{arbitrage equilibrium} is reached when the potential is maximized. We can determine the equilibrium utility, $h^*$, by maximizing the potential (see \cite{venkat2017book}), but there exists a simpler alternative that is more convenient for our purposes here. To analyze the equilibrium behavior, we can take the simpler agent-based perspective and exploit the fact that at equilibrium, all agents have the same effective utility, i.e. $h_i = h^*$, for all $i$. In other words, 

\begin{eqnarray}
        \alpha \rho^* - \beta \rho^{*2} + \ln (1-\rho^*) - \ln\rho^* = h^* 
        \label{eq:utility_equil}
\end{eqnarray}

\begin{figure}[ht]
\includegraphics[width=0.5\textwidth]{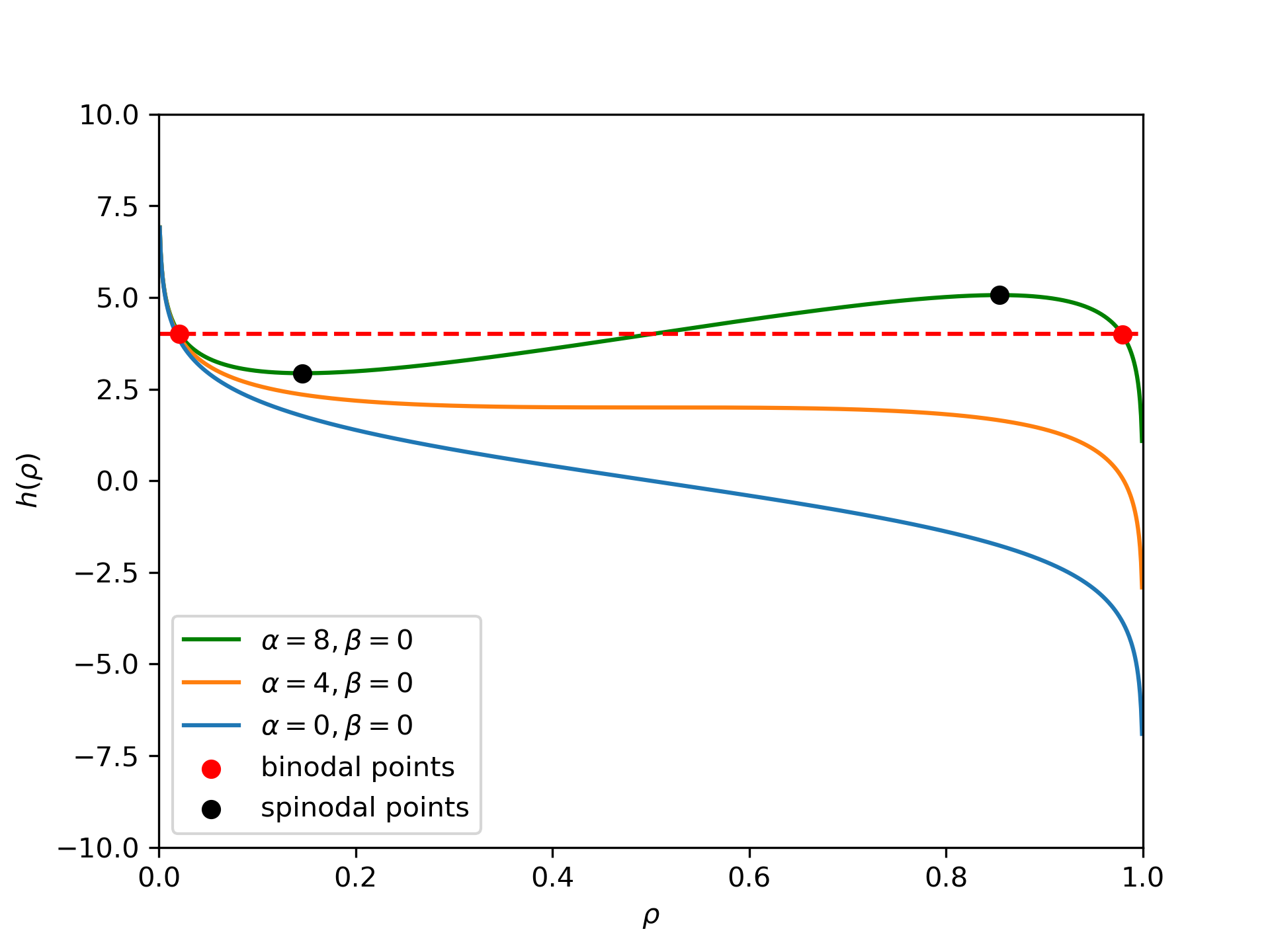}
\caption{Effective Utility vs Density: $h$ vs $\rho$ for different $\alpha$. The black points are the spinodal points ($\rho_{s1} = 0.146, h_{s1} = 2.934; \rho_{s2} = 0.854, h_{s2} = 5.066$). The red points are the binodal points ($\rho_{b1} = 0.021, h_{b1} = 4.00; \rho_{b2} = 0.979, h_{b2} = 4.00$).}
\label{fig:Utility_density_zero_beta}
\end{figure}

We explore numerically the behavior of $h^*$ as a function of $\rho ^*$ in ~\eqref{eq:utility_equil}, as shown in Fig.~\ref{fig:Utility_density_zero_beta} ( $\beta = 0$, different $\alpha$). Below a threshold value of $\alpha$ and $\beta$, the utility function is \textit{monotonic} and has a unique density (blue curve) for a given utility value. Above the threshold, the utility is non-monotonic (green curve) and can have multiple density values for the same utility. The red dotted line shows this. The orange curve is the threshold behavior. \\

Note that whether all agents remain in a single phase of uniform density dispersed throughout the region or separate into various groups is determined by the slope $\partial h/\partial \rho\big\vert_{\rho^*}$, which is the second derivative of $\phi$, $\partial^2 \phi /\partial^2 \rho\big\vert_{\rho^*}$. This behavior is mathematically equivalent to \textit{spinodal decomposition} in thermodynamics, widely studied, for example, in the phase separation of alloys and polymer blends~\cite{cahn1961spinodal, favvas2008spinodal}. \\ 

Although the mathematics of spinodal decomposition is equivalent to the segregation of social agents, there are significant differences in the mechanism of phase separation. In spinodal decomposition in a binary alloy formed with atoms of two metals, say $A$ and $B$, spinodal decomposition occurs when the change in free energy during the formation of $A-A$ bonds is less than that of the formation of $A-B$ bonds. However, our utility formulation allows phase separation even in systems with one-class agents. This is because our formulation introduces a disutility due to crowding and competition. These factors encourage agents to be part of large groups. The same behavior is also exhibited by systems with multiclass agents.\\

In thermodynamics, the phase between the spinodal points (discussed in Appendix A in more detail) is unstable as it corresponds to increasing the free energy of the system, and hence the single phase splits into two phases of different densities to lower the free energy.  For the same reason, the phases between the spinodal and binodal points are metastable, and the phases at the binodal points are stable. A similar behavior happens here in statistical teleodynamics as well (see Appendix A).\\

\section{Social Segregation in the one-class system: Parametric Regimes}

In this section, we determine the parametric regime in which social segregation occurs. To have social segregation in one-class systems, the utility curve should be non-monotonic in nature; more specifically, it should have a minimum and a maximum, as shown in Figure \ref{fig:Utility_density_zero_beta}. Differentiating Eq.\eqref{eq:Utility} we get

\begin{eqnarray}
    \frac{dh}{d \rho}= \alpha  - 2 \beta \rho -\frac{1}{\rho} -\frac{1}{(1-\rho)}
\end{eqnarray}

We know that the condition to have minimum or maximum is $\frac{dh}{d \rho} =0$. To have a minimum and a maximum, there should be two distinct real solutions for $\frac{dh}{d \rho} =0$ in the domain $0 \le \rho \le 1$.

\begin{eqnarray*}
    \frac{dh}{d \rho}= \alpha  - 2 \beta \rho -\frac{1}{\rho} -\frac{1}{(1-\rho)} =0
\end{eqnarray*}

\begin{eqnarray}
2 \beta \rho^3 - (\alpha + 2 \beta) \rho^2 + \alpha \rho -1 = 0
\label{eq:cubic_phase_separation_one_class}
\end{eqnarray}

\begin{figure}[ht]

\includegraphics[width=0.4\textwidth]{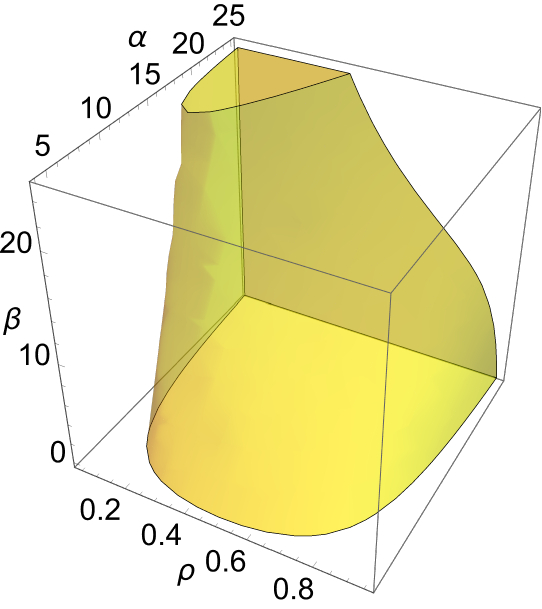}

\caption{The parametric space within the yellow region is where segregation is guaranteed to occur.}
\label{fig:phase_separation}

\end{figure}

\begin{figure}[ht]

\includegraphics[width=0.4\textwidth]{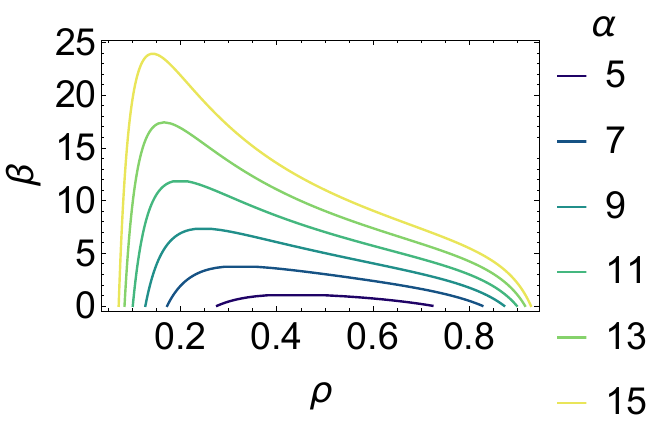}
\caption{This figure shows the cross-section of the solid from Figure \ref{fig:phase_separation}. Each line represents values of $\alpha$, showing the range of densities where segregation would occur for values of $\beta$.}
\label{fig:phase_separation_alpha_slice}

\end{figure}

This cubic equation can be solved numerically or using the Cardano formula. One can search in the $\alpha$-$\beta$  space for values that produce two real roots for Eq.\eqref{eq:cubic_phase_separation_one_class} in the domain $0 \le \rho \le 1$.  Fig.~\ref{fig:phase_separation}, shows the $\alpha$-$\beta$-$\rho_0$ region (shaded in yellow) within which phase separation \textit{occurs}. In Fig.~\ref{fig:phase_separation_alpha_slice}, we show the 2-D slices of the yellow region of spontaneous phase separation. For a given value of $\alpha$, $\beta$, and $\rho_0$, they show the loci of the two densities (i.e., the low and high density groups) of the corresponding equilibrium states of agents.

\section{Agent-based simulations: One-class system}

Agent-based simulations were performed for the parameters shown in Figure \ref{fig:Utility_density_zero_beta}. From the theory we had discussed earlier, we cannot expect phase separation to occur for $\alpha \le 4$ and $\beta=0$. However, phase separation can occur for $\alpha>4$ and $\beta=0$ when the initial average density is within the spinodal densities. Our agent-based simulations agree with this theoretical prediction. \\

\begin{figure*}[]
    \centering
    \includegraphics[width=\linewidth]{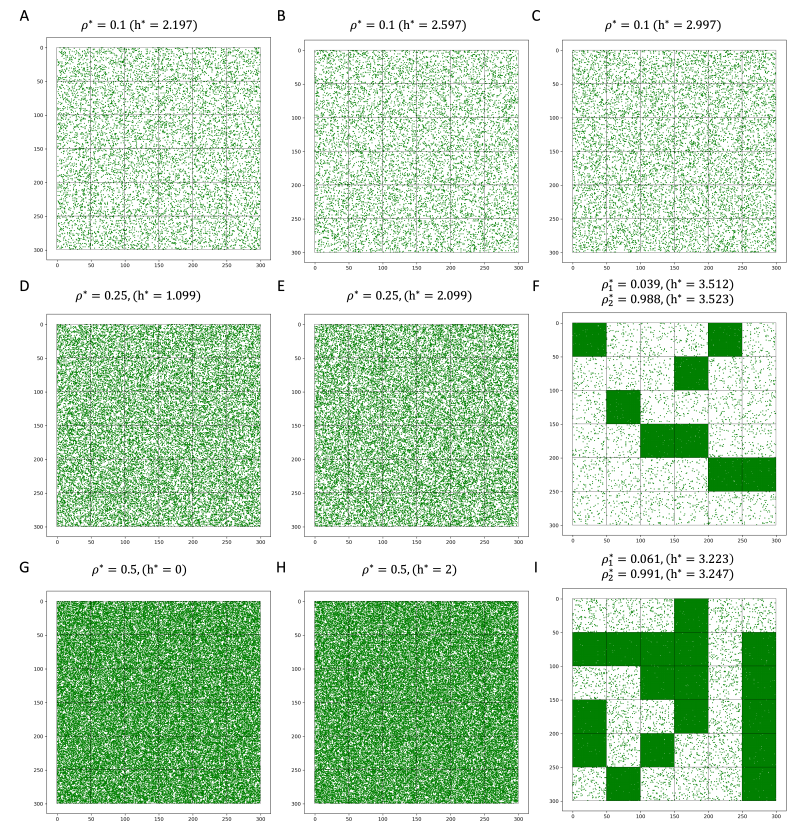}    
    \caption{ {\bf Equilibrium configuration after five million time steps on a $300\times 300$ grid, $M=50\times 50$ and $\beta=0$, for different $\rho_0$ and $\alpha$}. 
    Top row(\textit{A, B, C}), middle row(\textit{D, E, F}), and bottom row(\textit{G, H, I}) correspond to an average density of $\rho_0 = 0.1, 0.25, 0.5$, respectively. Left column (\textit{A, D, G}), middle column (\textit{B ,E, H}), and right column (\textit{C ,F, I}) correspond to $\alpha=0, 4, 8$, respectively. For single-phase systems, the final density is seen to be the same as $\rho_0$. For multiphase systems the density of the two phases are \textit{F.} 0.039 and 0.988 (utility of 3.512, 3.523, respectively) \textit{I.} 0.061 and 0.991 (utility of 3.223, 3.247 respectively)}
    \label{3X3}
\end{figure*}

The equilibrium configuration of the agents for nine sets of simulations is shown in Figure \ref{3X3}. The configurations shown in each row of Figure \ref{3X3} are for a constant density, and those shown in each column are for the same set of $\alpha$ and $\beta$. As expected, for $\alpha =0$ (configurations A, D, and G) and $\alpha =4$ (configurations B, E, and H), there is \textit{no} phase separation at any densities. For $\alpha =8$, phase separation \textit{does not} occur for $\rho_{0}=0.1$ (configuration C) because this density is outside the range of spontaneous decomposition. \\

However, at higher densities (configurations F and I), phase separation \textit{occurs} because the densities are now within the spinodal region (the yellow region in Fig.~\ref{fig:phase_separation}). In addition, note that the utility of both phases is equal (see Fig.~\ref{3X3} captions). Similar behavior was observed in cases where we kept $\alpha$ constant and varied $\beta$. \\

\begin{figure}[!ht]
\centering
\includegraphics[width=0.5\textwidth]{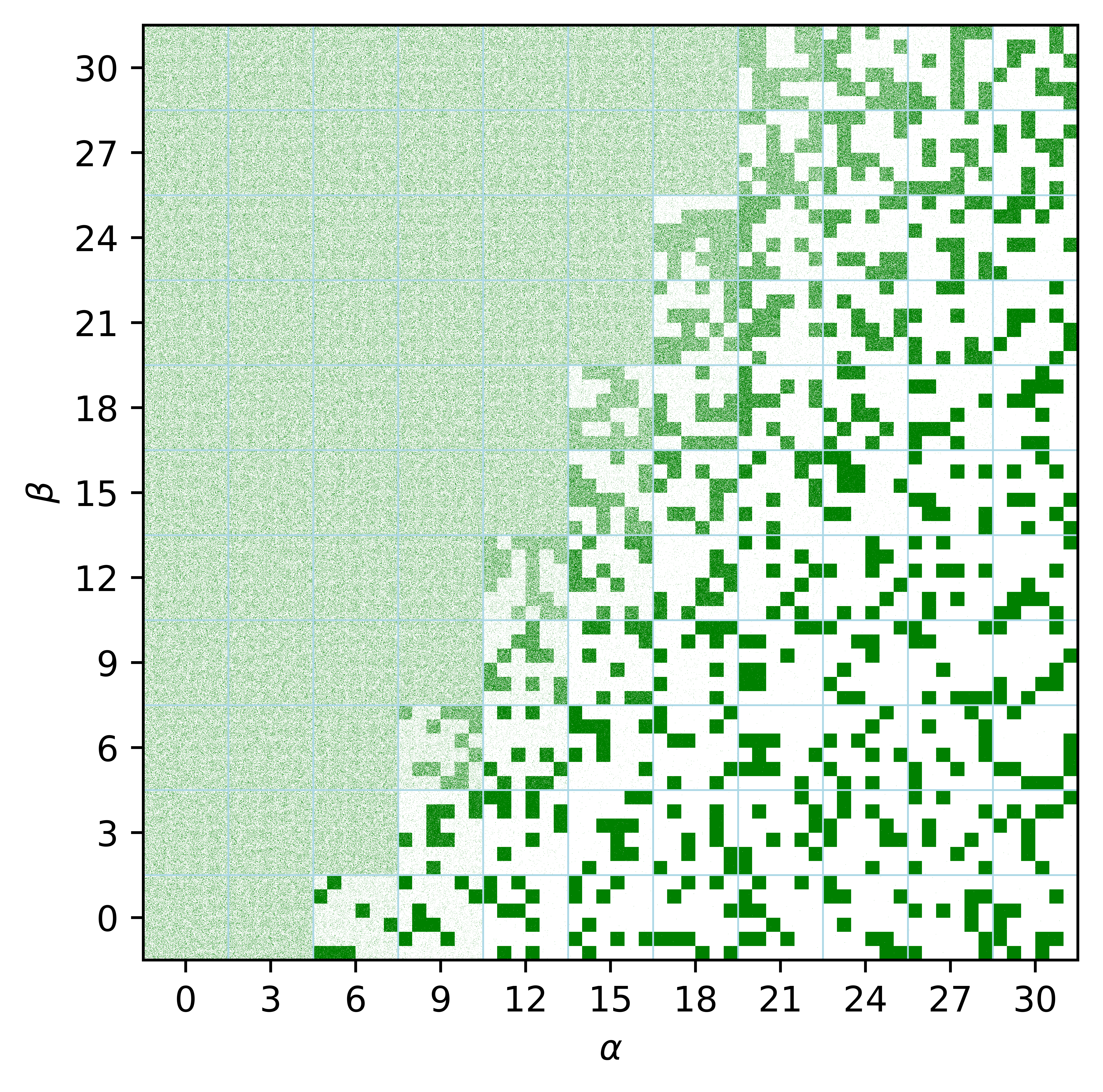}
\caption{Metaplot showing agent configurations for a wide range of $\alpha$ and $\beta$ ($N=22500$ for all simulations)}
\label{fig:one_class_metaplot}
\end{figure}

The parameter regime in which phase separation occurs was theoretically predicted and plotted in Figure \ref{fig:phase_separation}. Our numerical simulations agree with the theoretical predictions. The equilibrium agent configurations for a range of $\alpha$-$\beta$ values are shown in Figure \ref{fig:one_class_metaplot}. The $x$ and $y$ coordinates of each configuration indicate the values $\alpha$ and $\beta$ used in the simulation.\\

At low values of $\alpha$, there is no phase separation (for any $\beta$, see the uniform green region in Figure \ref{fig:one_class_metaplot}) because there is not much incentive to socially aggregate.  However, this changes as $\alpha$ increases, because the benefit of aggregation increases (see Eq.~\ref{eq:Utility}). So, social segregation occurs spontaneously, and the space divides itself into low-density (white squares) and high-density (green squares) blocks.\\

This behavior can also be explained by the monotonicity of the $h-\rho$ curve. At low values of $\alpha$ and high values of $\beta$, the $h-\rho$ curve is monotonically decreasing. For any fixed $\beta$, larger values of $\alpha$ result in nonmonotonic behavior for the $h-\rho$ curve. Similarly, for any fixed $\alpha$ ($\alpha > 4$), small values of $\beta$ lead to the non-monotonic behavior of the $h-\rho$ curve. This behavior can be observed in Figure \ref{fig:one_class_metaplot}. All the simulations shown in Figure \ref{fig:one_class_metaplot} are performed for N=22,500.\\

\section{Segregation in a Two-class system: Mathematical Analysis}

The formulation of a one-class system can readily be generalized to multiclass systems. Here, we discuss a two-class system, as an example, by modifying the vacancy terms in Eq. (\ref{eq:Utility_N_i}).
Consider a class of agents identified as \textit{green agents} and another class of agents identified as \textit{red agents}. Their utilities can be defined as follows.

\begin{eqnarray}
    \tilde{h}_{G,i}(N_{G,i},N_{R,i})&=& \tilde{\alpha}_G N_{G,i}  - \tilde{\beta}_G N_{G,i}^2 - \ln N_{G,i} \nonumber\\
    & & + \ln(M - N_{G,i}-N_{R,i})  \\ \nonumber\\
    \tilde{h}_{R,i}(N_{G,i},N_{R,i})&=& \tilde{\alpha}_R N_{R,i}  - \tilde{\beta}_R N_{R,i}^2 - \ln N_{R,i} \nonumber\\
    & & + \ln(M - N_{G,i}-N_{R,i}) 
    \label{eq:utility_two-class}
\end{eqnarray}

The utilities can be rewritten in terms of densities of the two classes, $\rho_G$ and $\rho_R$, as 

\begin{eqnarray}
    {h}_{G,i}(\rho_{G,i},\rho_{R,i})&=& \alpha_G \rho_{G,i}  - \beta_G \rho_{G,i}^2 - \ln \rho_{G,i} \nonumber \\
    & & + \ln(1 - \rho_{G,i}-\rho_{R,i})  
    \label{eq:utility_two-class_green}
\end{eqnarray}
\begin{eqnarray}
    {h}_{R,i}(\rho_{G,i},\rho_{R,i})&=& \alpha_R \rho_{R,i}  - \beta_R \rho_{R,i}^2 - \ln \rho_{R,i} \nonumber\\
    & & + \ln(1 - \rho_{G,i}-\rho_{R,i}) 
    \label{eq:utility_two-class_red}
\end{eqnarray}

where $\rho_{G,i}=\frac{N_{G,i}}{M}$ , $\rho_{R,i}=\frac{N_{R,i}}{M}$, $\alpha_G=\tilde{\alpha}_G M$, $\beta_G=\tilde{\beta}_G M^2$, $\alpha_R=\tilde{\alpha}_R M$,  and $\beta_R=\tilde{\beta}_R M^2$. (Refer to the supplementary material for the derivation.)\\

\begin{figure*}[]
\centering
\includegraphics[width=\linewidth]{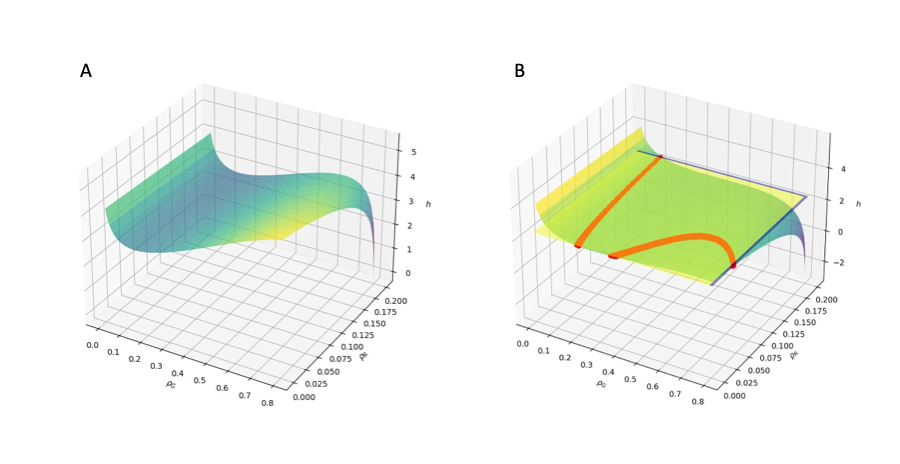}
\caption{A. $h$ vs. $\rho_G$, $\rho_R$  ($\alpha =5, \beta = 0$) B. Constant utility surface, h = 2.38 intersecting $h$ vs. $\rho_G$, $\rho_R$. The intersection points are marked in red.  ($\alpha =5, \beta = 0$)}
\label{fig:utility_density_2class}
\end{figure*}

For one-class systems, we showed that phase separation occurs when the utility-density curve is nonmonotonic. For two-class systems, the utility of a class depends not only on its density, but also on the density of the other class as well. The utility of each class in a two-class system is shown in Figure \ref{fig:utility_density_2class}A.  $\alpha=5$ and $\beta=0$ result in a nonconcave potential function. The nonconcavity of the potential is the primary requirement for phase separation. The conditions that result in non-concave potential are discussed in Appendix B.\\

In a one-class system, there were only three densities corresponding to utility in the phase-separation region. However, in two-class systems, infinite combinations of ($\rho_G, \rho_R$)  can exist for a specific utility value. These infinite density pairs are marked in red in Figure \ref{fig:utility_density_2class}B. However, the interaction of the two classes of agents restricts the number of possible combinations of coexisting densities. The detailed analysis is provided in Appendix B.  \\

\section{Agent-based simulations: two-class system}

Previously, we analyzed a one-class system and explained the qualitative changes in equilibrium configurations with changes in $\alpha$, $\beta$, and the initial densities. Similar analysis can be performed for the two-class system. \\

\begin{figure*}[]
    \centering
    \includegraphics[width=\linewidth]{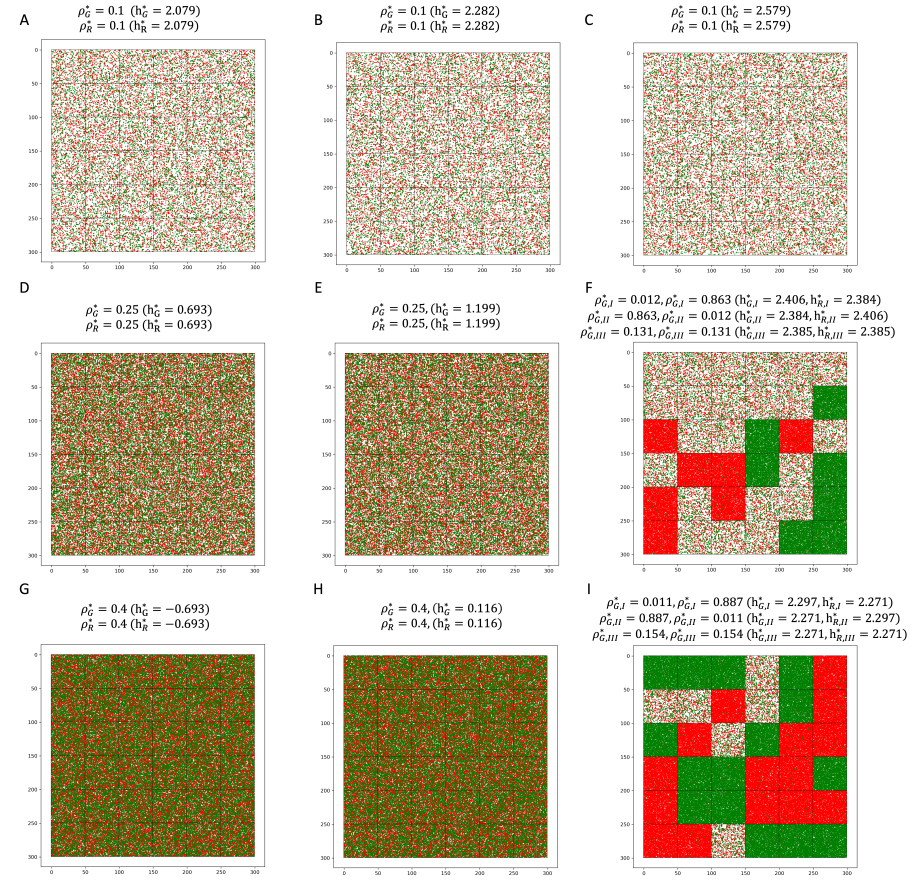}
    \caption{ {\bf Equilibrium configuration after five million time steps on a $300\times 300$ grid, $M=50\times 50$ and $\beta=0$, for different $\rho_0$ and $\alpha$}. 
    Top row(\textit{A, B, C}), middle row(\textit{D, E, F}), and bottom row(\textit{G, H, I}) correspond to an average density of $(\rho_{G,0}, \rho_{R,0}) = (0.1,0.1), (0.25,0.25)$, and $(0.4,0.4)$. Left column (\textit{A, D, G}), middle column (\textit{B, E, H}), and right column (\textit{C ,F, I}) correspond to $\alpha_G=\alpha_R=0, 2.023,$ and $ 5$, respectively.  For all simulations, $\beta_G=\beta_R=\beta=0$.  For single-phase systems, the final density is seen to be the same as ($\rho_{G0}, \rho_{R0}$). For multiphase systems, the densities of the two classes ($\rho^{*}_G,\rho^{*}_R$) in the three phases are F.  (0.012,0.863),(0.863, 0.012),(0.131,0.131) (with utilities (2.406, 2.384), (2.384, 2.406), (2.385, 2.385)) I. (0.011,0.887), (0.887,0.011), (0.154,0.154) (with utilities (2.297,2.271), (2.271,2.297), (2.271, 2.271))}
    \label{two_class_varying_alpha_3X3}
\end{figure*}

Consider Figure \ref{two_class_varying_alpha_3X3} showing equilibrium configurations of agents for different densities and $\alpha$s.
For simplicity, $\alpha_G$, and $\alpha_R$ are assumed to be equal (=$\alpha)$ and $\beta_G$ and $\beta_R$ are assumed to be equal (=$\beta$) in all the simulations reported in Figure \ref{two_class_varying_alpha_3X3}. $\beta$ is fixed to 0 in all simulations.  Each row of Figure \ref{two_class_varying_alpha_3X3} displays the equilibrium configurations for fixed densities of the two classes; $\rho_{G0}=\rho_{R0}=0.1$ in the upper row (A, B, C), $\rho_{G0}=\rho_{R0}=0.25$ in the middle row (D, E, F) and $\rho_{G0}=\rho_{R0}=0.4$ in the bottom row (G, H, I). Simulations were performed for three values of $\alpha$ ($\alpha=0, 2.023,5$). Each column in Figure \ref{two_class_varying_alpha_3X3} shows the equilibrium configuration for a fixed $\alpha$; $\alpha=0$ in the left column (A,D,G), $\alpha=2.023$ in the middle column (B, E, H), and $\alpha=5$ in the right column (C, F, I). \\

Like the one-class system, phase separation is not observed in the two-class system for small values of $\alpha$ ($\alpha=0$ and $2.023$) (Figure \ref{two_class_varying_alpha_3X3}A, B, D, E, G, H) for any densities. For $\alpha=5$, phase separation does not occur at $\rho_{G0}=\rho_{R0}=0.1$. \\

However, separation is observed at higher densities ($\rho_{G0}=\rho_{R0}=0.25$ and $\rho_{G0}=\rho_{R0}=0.4$). The nature of the game-theoretic potential can explain this behavior. Phase separation is observed when the potential-density surface is non-concave.  In Appendix B, we explain the conditions for the concavity of the potential. For concave functions, the eigenvalues of the Hessian will be non-positive. When at least one eigenvalue is positive, the potential is non-concave. 
The Hessian eigenvalues are provided in Table \ref{tab:hessian_eigenvalues} for each set of parameters reported in Figure \ref{two_class_varying_alpha_3X3}. It can be seen from Table \ref{tab:hessian_eigenvalues} that both Hessian eigenvalues (EV-1 and EV-2) are non-positive when $\alpha=0$ and $2.023$ and therefore the potential is concave for configurations A, B, D, E, G and H in Figure \ref{two_class_varying_alpha_3X3}. 

Theoretically, it can be verified that the potential is concave in the entire density domain when $\alpha \le 2.023$. The potential is globally non-concave when $\alpha > 2.023$ and $\beta=0$. For $\alpha=5$ and $\beta=0$, the potential is locally concave at the density (0.1,0.1) because, at this point, both eigenvalues of the Hessian are non-positive. Therefore, phase separation is not observed at this set of densities. However, one of the eigenvalues is positive at densities (0.25,0.25) and (0.4,0.4) (EV-1 $=1>0$ for configuration F and EV-1$=2.5>0$ for configuration I). Therefore, the potential is non-concave, and as a consequence, phase separation is observed in configurations F and I. 

\begin{table*}[!ht]
    \centering
    \caption{Eigenvalues of the Hessian and local concavity for configurations reported in Figure \ref{two_class_varying_alpha_3X3}.}
    \begin{tabular}{|m{2.0cm}|m{1.0cm}|m{1.0cm}|m{2.0cm}|m{2.0cm}|m{2.0cm}|m{2.0cm}|m{2.0cm}|}
        \hline
         Configuration & $\alpha$& $\beta$ & $\rho_{G0}$  &  $\rho_{R0}$ & EV-1 & EV-2 & Local \\
    in Figure \ref{two_class_varying_alpha_3X3}& &  &  &   &  &  & concavity \\
         
         \\ [-1em]
         \hline
    A & 0 & 0 & 0.1 & 0.1 & -10 & -12.5 & Concave\\
    B & 2.023 & 0 & 0.1 & 0.1 & -7.977 & -10.477 & Concave\\
    C & 5 & 0 & 0.1 & 0.1 & -5. & -7.5 & Concave \\
    D & 0 & 0 &0.25 & 0.25 &-4 & -8 & Concave\\
    E & 2.023 & 0 & 0.25 & 0.25 & -1.977 &-5.977 & Concave\\
    F & 5 & 0 & 0.25 & 0.25 & 1 & -3 & Non-concave\\
    G & 0 & 0 &0.4 & 0.4 &-2.5 & -12.5 & Concave\\
    H & 2.023 & 0 & 0.4 & 0.4 &-0.477 &-10.477 & Concave\\
    I & 5 & 0 & 0.4 & 0.4 & 2.5 &-7.5& Non-concave\\
    \hline  
    \end{tabular}
      \label{tab:hessian_eigenvalues}
    \end{table*}
    
\begin{figure}[]
\centering
\includegraphics[width=0.5\textwidth]{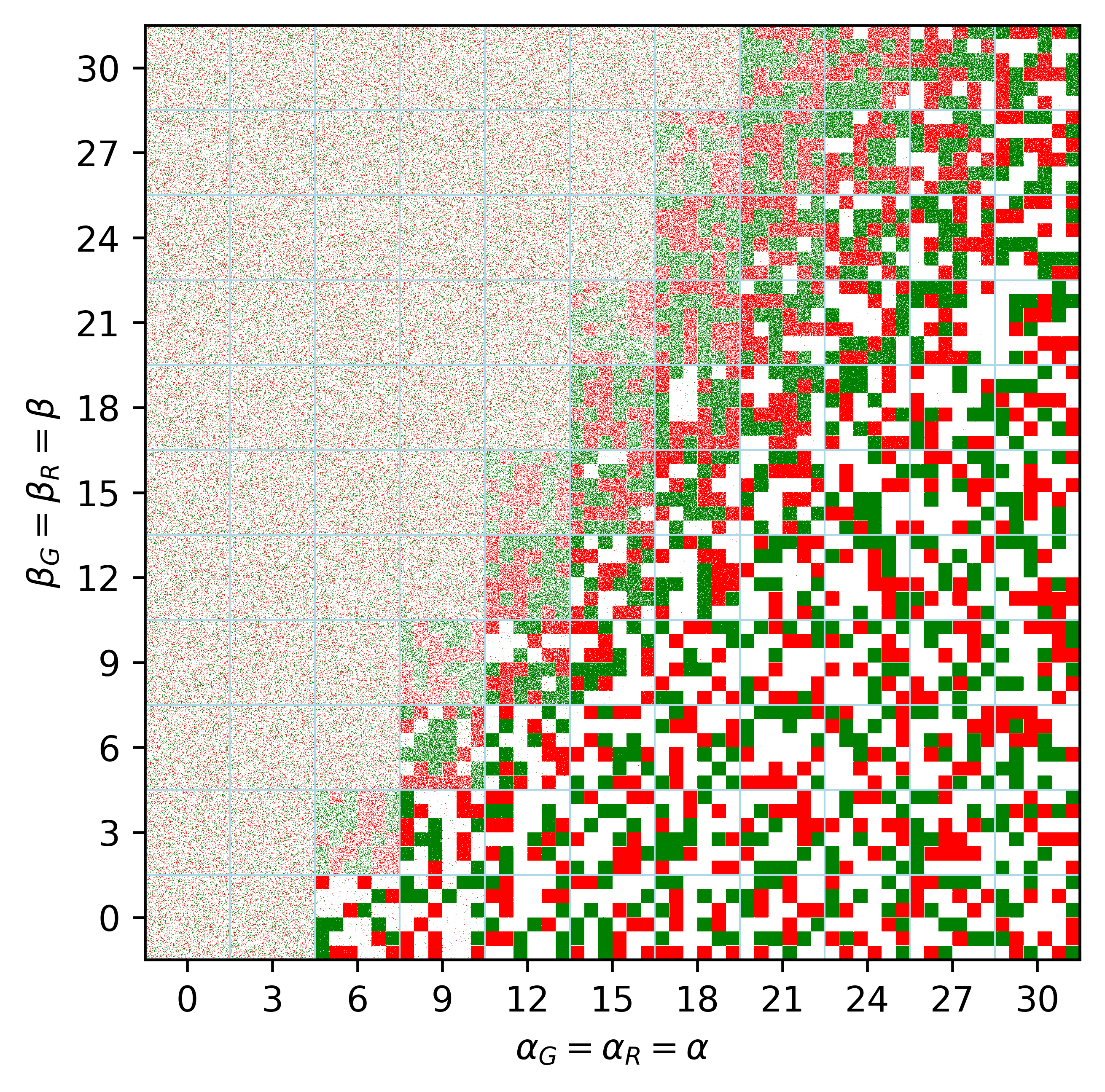}
\caption{ A metaplot showing agent configurations for a wide range of $\alpha$ and $\beta$ ($\alpha_G=\alpha_R=\alpha$ and $\beta_G=\beta_R=\beta$). For all simulations, $N_G=N_R=22500$.}
\label{fig:two_class_metaplot}
\end{figure}

The parameter regime where phase separation occurs was theoretically predicted and plotted in Figure \ref{fig:two_class_phase_separation_alpha_beta_regime}. Our numerical simulations agree with the theoretical predictions. The equilibrium agent configurations for a range of $\alpha$-$\beta$ values are shown in Figure \ref{fig:two_class_metaplot}. The $x$ and $y$ coordinates of each configuration indicate the values $\alpha$ and $\beta$ used in the simulation. At low values of $\alpha$, there is no phase separation because there is not much incentive to get together.  As $\alpha$ increases, the agents come together. For any $\alpha$, the disutility due to crowding increases as $\beta$ increases. Mathematically, at low values of $\alpha$ and high values of $\beta$, the potential $\phi$ is concave (characterised by non-positive eigenvalues). For any fixed $\beta$, larger values of $\alpha$ results in non-concave potential. Similarly, for any fixed $\alpha$ ($\alpha>2.023$), small values of $\beta$ leads to the non-concave potential. Therefore, phase separation is expected at large values of $\alpha$ for a specific $\beta$ and at small values of $\beta$ for a specific $\alpha$. This behavior is observed in our simulations reported in Figure \ref{fig:two_class_metaplot}. All the simulations shown in Figure \ref{fig:two_class_metaplot} performed for $N_G=N_R=22,500$.

\section{Methodology-Python Simulations}
We developed Schelling-like models for one-class and two-class systems in the Python environment. The inputs to the program are the number of agents of each type, $\alpha$s and $\beta$s. The number of agents in each class remains fixed through the simulation and is given by $\Bar{\rho_{G0}} \times Q \times M$ for the green agents and $\Bar{\rho_{R0}} \times Q \times M$ for the red agents. At each iteration, an agent and a vacant coordinate outside of the agent's neighborhood are chosen at random. The agent's utility in the current position is compared to the agent's utility if the agent were to hypothetically move to the new vacant coordinate. If the agent's utility is increased by moving to the new location, the agent moves and the respective density and utility values are updated. If not, the agent stays in its original position. Five million such offers are made to the agents. For the appropriate parameter set, the movement of the agents results in phase separation.\\

\section{Discussion}
In this paper, we have presented a mathematical analysis of the decision dynamics of socioeconomic segregation in one- and two-class systems using a utility-driven game-theoretic model. For one-class systems, we showed both analytically and in simulations that the non-monotonic structure of the utility model causes social segregation. Depending on the parameters $\alpha$ and $\beta$, this usually occurs at higher densities. One can also predict and explain this segregation from the perspective of a system trying to maximize its game-theoretic potential. We have connected these two ideas by showing that the non-monotonic nature of the utility function results in non-concave potentials. The results of our agent-based simulations support these theoretical predictions. The parametric regime of segregation was identified using theory and confirmed by agent-based simulations. \\

More importantly, these predictions align with the results reported in the study by Nilforoshan et al.\cite{nilforoshan2023human}, where they observed social segregation in larger, denser cities. As noted, they observe that "The consistent result that larger, denser cities are more segregated runs counter to the hypothesis that such cities promote socioeconomic mixing by attracting diverse individuals and constraining space in ways that oblige them to encounter one
other." Our theory predicts this density-dependent segregation, as we discussed above. \\

They further state, "Our results support the opposite hypothesis: big cities allow their inhabitants to seek out people who are more like themselves." This is also expected in our theory, as seen in Fig.~\ref{fig:one_class_metaplot}.  As $\alpha$ increases, the system tends to segregate. The parameter $\alpha$ measures the social utility that an agent enjoys by being with others. In larger cities, as Nilforoshan et al.\cite{nilforoshan2023human} study observes, there are many social options. This implies that $\alpha$ will be higher in larger cities than in smaller ones. Therefore,  as seen in Fig.~\ref{fig:one_class_metaplot}, it is not surprising that the population segregates into different groups. In fact, this is expected by our model. \\

For the two-class system, as in the one-class system, social segregation occurs when the potential is non-concave. The negative-definiteness of the Hessian of the game-theoretic potential determines the concavity. Our simulations of two-class systems agree with the theoretical predictions. We have also determined the parametric regime of phase separation for the two-class systems.  \\

Finally, we wish to direct the reader's attention to an interesting observation. We note that the effective utility ($h^*$) enjoyed by the agents in the low- and high-density phases is the \textit{same} (see Fig.~\ref{3X3}F and I), as expected, because this is an arbitrage equilibrium. From a socioeconomic perspective, this is an interesting result. Interpreting the effective utility as a measure of "happiness" or "satisfaction", we see that the two different socioeconomic groups are equally "happy" in their respective environments. Even though they are segregated, they are both equally "satisfied" with their lifestyles in our ideal society. To put this a bit more colorfully, the person enjoying a beverage with a small group of friends in a low-density small town is just as "happy" as his/her larger city counterpart in a fancy and crowded restaurant in our ideal society. This harmonious outcome, despite segregation, is not necessarily bad. However, since "happiness" is such an elusive concept, we wish to emphasize and caution that this ideal harmonious outcome, as envisioned in our model, might not occur in real-world societies. Although segregated populations enjoy the same effective utility in our ideal society, this might not be the case in the real world due to various social, economic, and political policies and constraints.  \\

\pagebreak
\appendix
\section{Stability analysis of one-class systems}

In Fig~\ref{fig:Utility_density_zero_beta}, we observe that for $\alpha = 0$ (blue curve) and $\alpha = 4$ (orange curve), $\partial h/\partial \rho \le 0$ (i.e. negative slope; recall that $\partial h/\partial \rho $ = $\partial^2 \phi /\partial^2 \rho $). In such a parameter regime, phase separation does not occur. However, for higher values of $\alpha$, regions with $\partial h/\partial \rho>0$ (i.e., positive slope) phase separation develop. 
\\

\begin{figure}[]
\includegraphics[width=0.5\textwidth]{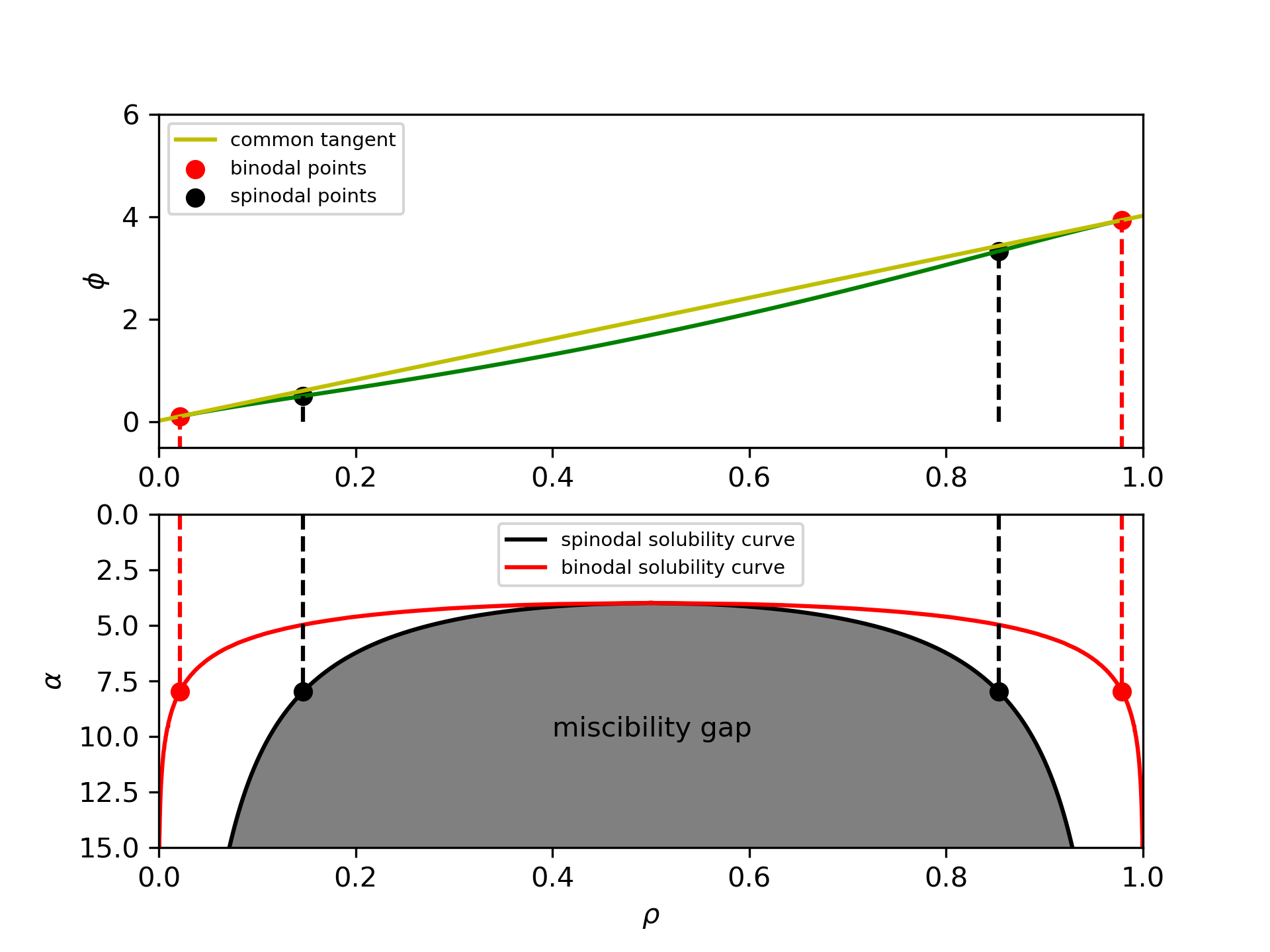}
\caption{Game potential ($\phi$) curve and the spinodal and binodal points. For $\alpha =8, \beta = 0$, the spinodal densities are $0.146$ and $0.854$; the binodal densities are $0.021$ and $0.979$.}
\label{fig:SpinodalBinodal}
\end{figure}

We better understand this from Fig.~\ref{fig:SpinodalBinodal}. The upper part of this figure shows the potential ($\phi$) vs the density ($\rho$) curve (in green) for $\alpha =6$, $\beta =0$. 
The plotted equation is \\

$\phi = \alpha \frac{\rho^2}{2} - \beta \frac{\rho^3}{3}- \rho \ln{\rho}- (1-\rho) \ln{(1-\rho)}$ \\

Spinodal points are shown as black dots, where $\partial h/\partial \rho\big\vert_{\rho^*}$ = $\partial^2 \phi /\partial^2 \rho\big\vert_{\rho^*} = 0$. The corresponding spinodal points are also shown in Fig.~\ref{fig:Utility_density_zero_beta} as black dots on the green curve ($\alpha = 8$, $\beta = 0$). Fig.~\ref{fig:SpinodalBinodal} also shows the binodal points (in red, connected by the common tangent line), where $\partial h/\partial \rho\big\vert_{\rho^*}$ = $\partial^2 \phi /\partial^2 \rho\big\vert_{\rho^*} < 0$. The corresponding binodal points are seen in Fig.~\ref{fig:Utility_density_zero_beta} as red points connected by the red dotted line. As we see, the two binodal points enjoy the same utility (4.00), which is the arbitrage equilibrium.\\

The lower part of Fig.~\ref{fig:SpinodalBinodal} shows the loci of binodal points (red curve) and spinodal points (black curve) for different values of $\alpha$ ($\beta = 0$). As $\alpha$ changes, the binodal and spinodal points change, and for $\alpha > 4$ ($\beta = 0$) they disappear. Within the spinodal region, shown in dark gray, known as the miscibility gap in thermodynamics, a single phase of uniform density is unstable and would split into two phases of different densities. The reason is that the potential $\phi$ of a large group here is less than the sum of the two potentials of the low-density group and the high-density group at the binodal points. \\

We see this geometrically from the common tangent line connecting the binodal points, which is above the single-phase green curve between the spinodal points. Agents in such regions will be self-driven towards the high-density binodal point to increase their utility. Therefore, $\phi$ increases and the system splits into two groups of different densities. \\

Thus, for the green curve in Fig.~\ref{fig:Utility_density_zero_beta}, a self-organized, utility-driven, stable phase separation occurs spontaneously at the binodal points (red dotted line) at the arbitrage equilibrium. Although the miscibility gap is \textit{unstable}, the region immediately outside it, between the black and red curves, is \textit{metastable}. Beyond the red curve, one has a stable single phase of uniform density - no phase separation here.  \\

In summary, for high values of $\alpha$ (e.g., green curve in Fig.~\ref{fig:Utility_density_zero_beta}), combined with average densities in the miscibility gap, we observe the spontaneous emergence of two phases, high- and low-density groups of agents, at arbitrage equilibrium, socially driven by the self-actuated pursuit of maximum utility by the agents. \\

Intuitively, in the high-density phase, agents derive so much more benefit from the affinity term (due to the high $\alpha$) that it more than compensates for the disutilities due to congestion and competition, thus yielding a high effective utility. Similarly, in the low-density phase, the benefits of reduced congestion and lower competition combined with increased option benefit more than compensate for the loss of utility from the affinity term. \\

Thus, every agent enjoys the same effective utility $h^*$ in one phase or the other at equilibrium. This causes equilibrium because, as noted, there is no more arbitrage incentive left for agents to switch neighborhoods. \\

As noted above, this analysis is mathematically equivalent to spinodal decomposition in statistical thermodynamics, with an important difference. In statistical thermodynamics, agents try to \textit{minimize} their chemical potentials and the free energy of the system. Here, in statistical teleodynamics, agents try to \textit{maximize} their utilities ($h_i$) and the game-theoretic potential ($\phi$). In thermodynamics, chemical potentials are equal at phase equilibrium. In teleodynamics, the utilities are equal at the arbitrage equilibrium. The parallel is striking, but not surprising, because, as Venkatasubramanian has shown~\cite{venkat2017book, venkat2022unified}, statistical teleodynamics is the generalization of statistical thermodynamics for goal-driven agents.\\

\section{Mathematical analysis of two-class systems}

Consider the case, $\alpha_G=\alpha_R=\alpha=5$ and $\beta_G = \beta_R =\beta =0$. For now, assume that the equilibrium utility is $ h^{*}=2.38$ (this is the equilibrium utility observed in the agent-based simulations). Due to the symmetry ($\alpha_G= \alpha_R$ and $\beta_G = \beta_R$), both classes of agents are expected to exhibit identical behavior. This implies that if there exists a \textit{green-rich} phase, referred to as \textit{Phase-I}, with densities $\rho^{*}_{G, I}$ and $\rho^{*}_{R, I}$, then there must also exist a \textit{red-rich} phase, referred to as \textit{Phase-II} with densities $\rho^{*}_{G, II}$ and $\rho^{*}_{R, II}$ such that $\rho^{*}_{G, I}= \rho^{*}_{R, II}$ and $\rho^{*}_{R, I}=\rho^{*}_{G, II}$. \\

In addition to these phases, there may also exist other phases where both classes have the same densities. At equilibrium, the utilities of the two classes in each phase should be equal owing to the symmetry of the parameters, i.e.,

\begin{widetext}
    \begin{eqnarray}
    h^{*}(\rho^{*}_{G,i},\rho^{*}_{R,i})= \alpha_G \rho^{*}_{G,i} - \beta_G  {\rho^{*}}^2_{G,i}- \ln {\rho^{*}}_{G,i} + \ln(1-{\rho^{*}}_{G,i}-{\rho^{*}}_{R,i})
    = \alpha_R \rho^{*}_{R,i} - \beta_R  {\rho^{*}}^2_{R,i}- \ln {\rho^{*}}_{R,i} + \ln(1-{\rho^{*}}_{G,i}-{\rho^{*}}_{R,i})  \nonumber \\
    \label{eq:equality_of_utility_1}
\end{eqnarray}
\end{widetext}

The constraint on the equilibrium densities emerging from Equation \eqref{eq:equality_of_utility_1} is 

\begin{eqnarray}
    \alpha \rho^{*}_{G,i} - \beta  {\rho^{*}}^2_{G,i}- \ln {\rho^{*}}_{G,i} 
    =\alpha \rho^{*}_{R,i} - \beta  {\rho^{*}}^2_{R,i}- \ln {\rho^{*}}_{R,i} \nonumber \\
    \label{eq:equality_of_utility_2}
\end{eqnarray}

\begin{figure*}[!ht]
\centering
\includegraphics[width=\linewidth]{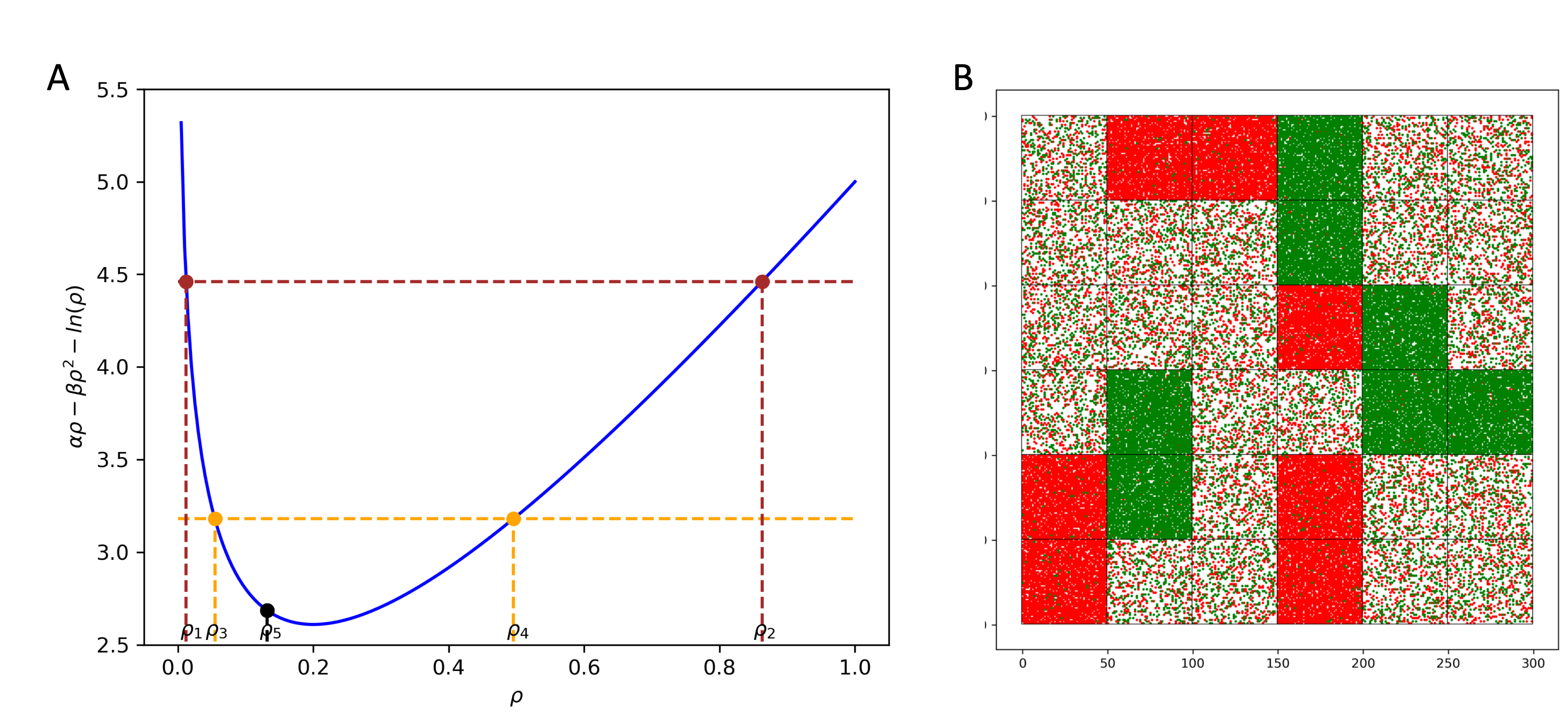}
\caption{A. $g(\rho)=\alpha \rho - \beta  {\rho}^2- \ln {\rho}$ vs. $\rho$ ($\alpha =5, \beta = 0$). The marked are solutions that satisfy Eq.\eqref{eq:equality_of_utility_1}, i.e., all  combinations ($\rho_1,\rho_2$), ($\rho_2,\rho_1$),($\rho_3,\rho_4$), ($\rho_4,\rho_3$) and ($\rho_5,\rho_5$) result in the same utility value, $ h^{*}=2.38$. $\rho_1=0.012,\rho_2= 0.863, \rho_3 = 0.055, \rho_4 = 0.495, \rho_5=0.132$  B. Equilibrium configuration attained in the agent-based simulations for $\alpha_G=\alpha_R=5, \beta_G=\beta_R=0, N_G=N_R=22,500$ at the end of 5 million iterations.} 
\label{fig:utility3terms}
\end{figure*}

Let us determine the densities satisfying Eq.\eqref{eq:equality_of_utility_1}. Refer to the plot of $g (\rho)=\alpha \rho - \beta  {\rho}^2- \ln {\rho}$ shown in Figure \ref{fig:utility3terms} A. It can be observed that for a range of values of $g$, there exist two densities ($\rho$) with the same value of $g$. Two such solutions are shown with brown and orange points. The two densities represented by brown points have the same value of $g$. Similarly, the two densities represented by the orange points have the same value of $g$.  Among the two brown points, let $\rho_1$ be the density of green agents, then $\rho_2$ will be the density of red agents in that phase (\textit{red-rich} phase). Following the symmetry argument, there exists another phase (\textit{green-rich} phase) with density $\rho_1$ for the red agents and density $\rho_2$ for the green agents. If any such pair of densities results in the equilibrium utility, $h^{*}$, then that can be considered as a potential candidate for the densities in one of the phases. \\

Another possibility is the existence of other phases where both classes have equal density. An example of such a pair of densities is represented by the black point. \\

\begin{table*}[!ht]
    \centering
    \caption{Derivative of the utilities of the two classes at densities reported in Figure \ref{fig:utility3terms}.}
    \begin{tabular}{|m{2.0cm}|m{1.0cm}|m{1.0cm}|m{1.0cm}|m{2.0cm}|m{2.0cm}|m{2.0cm}|m{2.0cm}|}
        \hline
         Candidate Phase&  $\rho_G$  &  $\rho_R$ &   h &  $\frac{\partial{h_G}}{\partial{\rho_G}}$& $\frac{\partial{h_G}}{\partial{\rho_R}}$ &  $\frac{\partial{h_R}}{\partial{\rho_G}}$& $\frac{\partial{h_R}}{\partial{\rho_R}}$\\ 
         \\ [-1em]
         \hline
        I&0.012 &  0.863  & 2.38 & -84.409&-7.995 &-7.995 &-4.154 \\
        II&0.863 &  0.012  & 2.38 & -4.154 &-7.995 & -7.995&-84.409\\
        III&0132 & 0.132    & 2.38 & -3.951 &-1.358& -1.358 & -3.951 \\
        IV&0.055 &   0.495 & 2.38 &  -15.512&-2.223& -2.223& 0.759\\
        V&0.495&  0.055   & 2.38 &0.759  &-2.223& -2.223&-15.511\\
        \hline
    \end{tabular}
    
    \label{tab:utility_derivatives}
\end{table*}

All densities marked in Figure \ref{fig:utility3terms}A are solutions to Eq.\eqref{eq:equality_of_utility_1}, i.e., the density combinations ($\rho_1,\rho_2$), ($\rho_2,\rho_1$),($\rho_3,\rho_4$), ($\rho_4,\rho_3$) and ($\rho_5,\rho_5$) result in the same utility value, $ h^{*}=2.38$. Therefore, we have five potential phases into which the mixture may separate. Our extensive search in the density domain confirmed that there is no other solution to Eq. \eqref{eq:equality_of_utility_1}. \\

Can all these five phases coexist? Any phase with $\frac{\partial h_G}{\partial \rho_{G_i}}>0$, $\frac{\partial h_G}{\partial \rho_{R_i}}>0$, $\frac{\partial h_R}{\partial \rho_{G_i}}>0$ or $\frac{\partial h_R}{\partial \rho_{R_i}}>0$ is unstable. These derivatives evaluated in the five density pairs are provided in Table \ref{tab:utility_derivatives}. Note that for Phases IV and V, one of the utility derivatives is positive. For Phase IV, $\frac{\partial{h_R}}{\partial{\rho_R}}>0$, indicating that the red agents of other phases can move to Phase IV and increase their utility.  These movements will increase the density of the red agents in Phase IV, and therefore Phase IV is unstable. Similar arguments can be applied to the green agents in Phase V. Therefore, for a phase to coexist with other phases, the utility derivatives must be non-positive. In summary, at equilibrium, we anticipate phases I, II, and III to coexist.\\

In addition to the above requirements, considering finite number of blocks and the definition of density of each class, the candidate density pairs must also satisfy the constraint that the number of blocks belonging to each phase must be an integer. Consider the separation of a mixture into three phases. Let the densities of the two classes in the three phases be represented as $\rho^{*}_{G,I}$,$\rho^{*}_{R,I}$,$\rho^{*}_{G,II}$,$\rho^{*}_{R,II}$, $\rho^{*}_{G,III}$ and $\rho^{*}_{R,III}$. The corresponding number of blocks in each phase is represented by $n_I$,$n_{II}$ and $n_{III}$. Then the conservation of the number of agents requires the following constraints to be satisfied. 

\begin{eqnarray}
    M \left(n_I \rho^{*}_{G,I} + n_{II}  \rho^{*}_{G,II}+ n_{III}   \rho^{*}_{G,II} \right)
    &=& N_G \\
    M (n_I  \rho^{*}_{R,I} + n_{II}  \rho^{*}_{R,II}+ n_{III}   \rho^{*}_{R,II})
    = N_R \\
    n_I +n_{II} +n_{III}=Q
    \label{density_balance}
\end{eqnarray}
where $n_I, n_{II}, n_{III} \in \{0,1,2...,Q\}$. These are soft constraints, meaning that minor differences in the calculated densities are expected in the simulations to ensure $n_I$,$n_{II}$ and $n_{III}$ are integers.\\

The equilibrium configuration resulting from the simulation for $\alpha=5$ and $\beta=0$ is provided in Figure \ref{fig:utility3terms}B. As predicted by the theory, the simulation results in three phases. Moreover, the densities of the two classes in each phase observed in the simulation are exactly the same as those of the predictions. \\

\subsection*{Game-theoretic potential: Two-class system}
The utilities of the two classes provided in Eq. \eqref{eq:utility_two-class_green} and \eqref{eq:utility_two-class_red} can be used to compute the game-theoretic potential of the system as shown in \cite{sandholm2010population}. The potential is given by

\begin{eqnarray}
    \phi
    &=\sum_{i=1}^{Q} \Bigg( \alpha_G \frac{\rho_{G,i}^2}{2} - \beta_G  \frac{\rho_{G,i}^3}{3}-\rho_{G,i} \ \ln \rho_{G,i} \nonumber \\
    & + \alpha_R \frac{\rho_{R,i}^2}{2} - \beta_R \frac{\rho_{R,i}^3}{3}  -\rho_{R,i} \ \ln \rho_{R,i} \nonumber \\
    & - (1-\rho_{G,i}-\rho_{R,i})\ln(1-\rho_{G,i}-\rho_{R,i}) \Bigg) \label{eq:potential}
\end{eqnarray}

\subsection*{Parametric regime: Two-class system}
For the one-class system, the parametric regime for phase separation was calculated from the derivative of the utility curve. For two-class systems, there are two independent densities, and therefore, one needs to consider the derivatives of utility with respect to both densities. In other words, one needs to analyze the Hessian of the potential to predict phase separation. Phase separation does not occur when the potential curve is purely concave. Phase separation can be expected for parameters that make the potential non-concave.

The Hessian of the potential is defined as:
\begin{equation}
H=\begin{bmatrix}
\frac{\partial^2 \phi}{\partial {\rho_G}^2} & \frac{\partial^2 \phi}{\partial {\rho_G} \partial {\rho_R}} \\ \\[-1em]
\frac{\partial^2 \phi}{\partial {\rho_R} \partial {\rho_G}}&  \frac{\partial^2 \phi}{\partial {\rho_R}^2} 
\end{bmatrix}
\label{hessian}
\end{equation}

Substituting the potential expression, we get, 
\begin{widetext}
    \begin{equation}
    H=\begin{bmatrix}
    \alpha_G - 2\beta_G \rho_G -\frac{1}{\rho_G} -\frac{1}{1- \rho_G - \rho_R} & \frac{-1}{1- \rho_G - \rho_R} \\ \\
    \frac{-1}{1- \rho_G - \rho_R}&  \alpha_R - 2\beta_R \rho_R -\frac{1}{\rho_R} -\frac{1}{1- \rho_G - \rho_R}
    \end{bmatrix}
    \label{hessian2}
\end{equation}
\end{widetext} 

\begin{figure}[ht]
\includegraphics[width=0.5\textwidth]{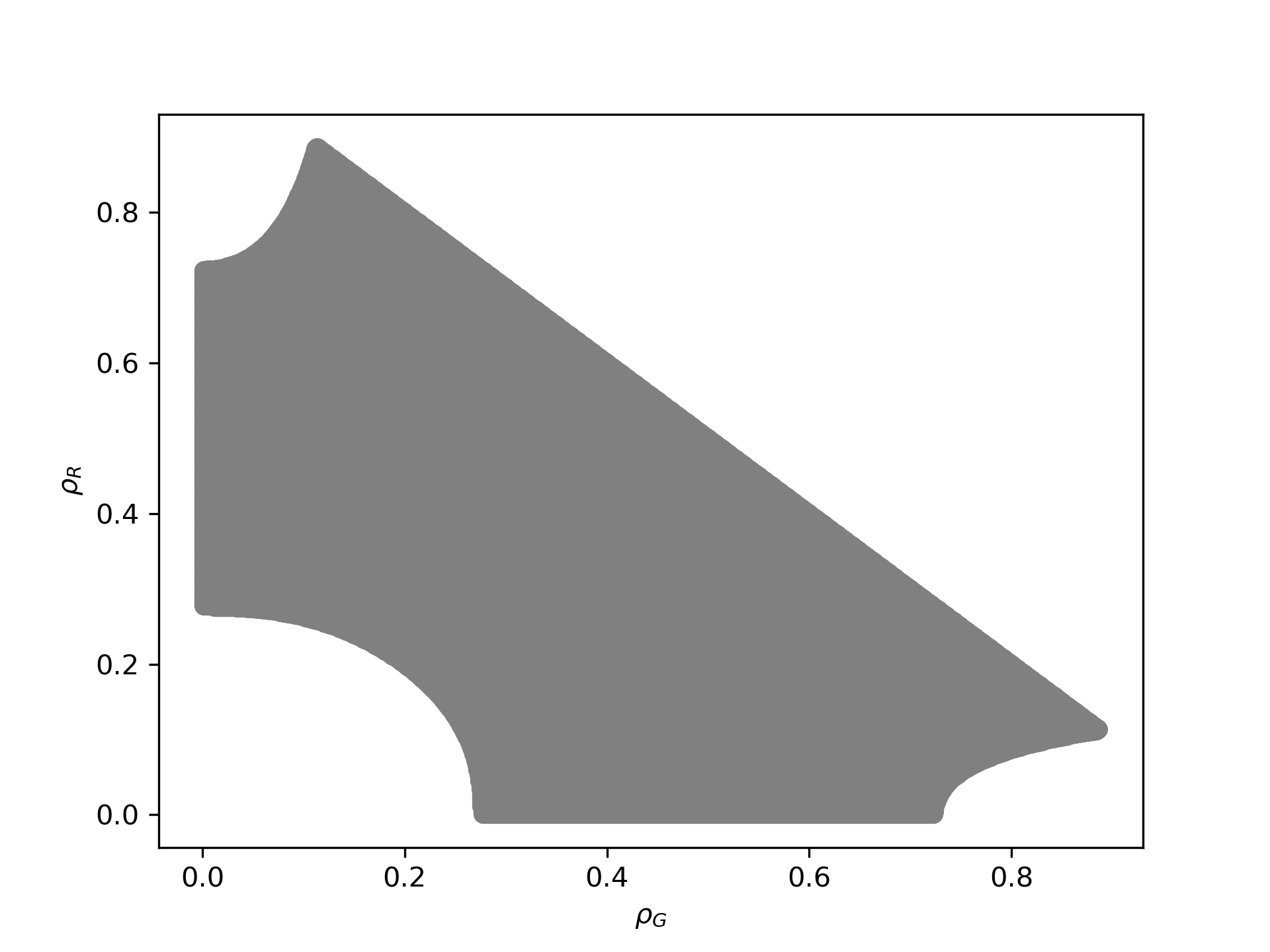}
\caption{Shaded region indicates the region of densities of the two phases where at least one eigenvalue of the Hessian matrix is positive ($\alpha_G=\alpha_R=\alpha=5$ and $\beta_G=\beta_R=\beta=0$)}. Phase separation occurs when the density of the initial phase is in the shaded region.
\label{fig:two_class_phase_separation_alpha_5_beta_0_density_regime}
\end{figure}  

\begin{figure}[h!]
\includegraphics[width=0.5\textwidth]{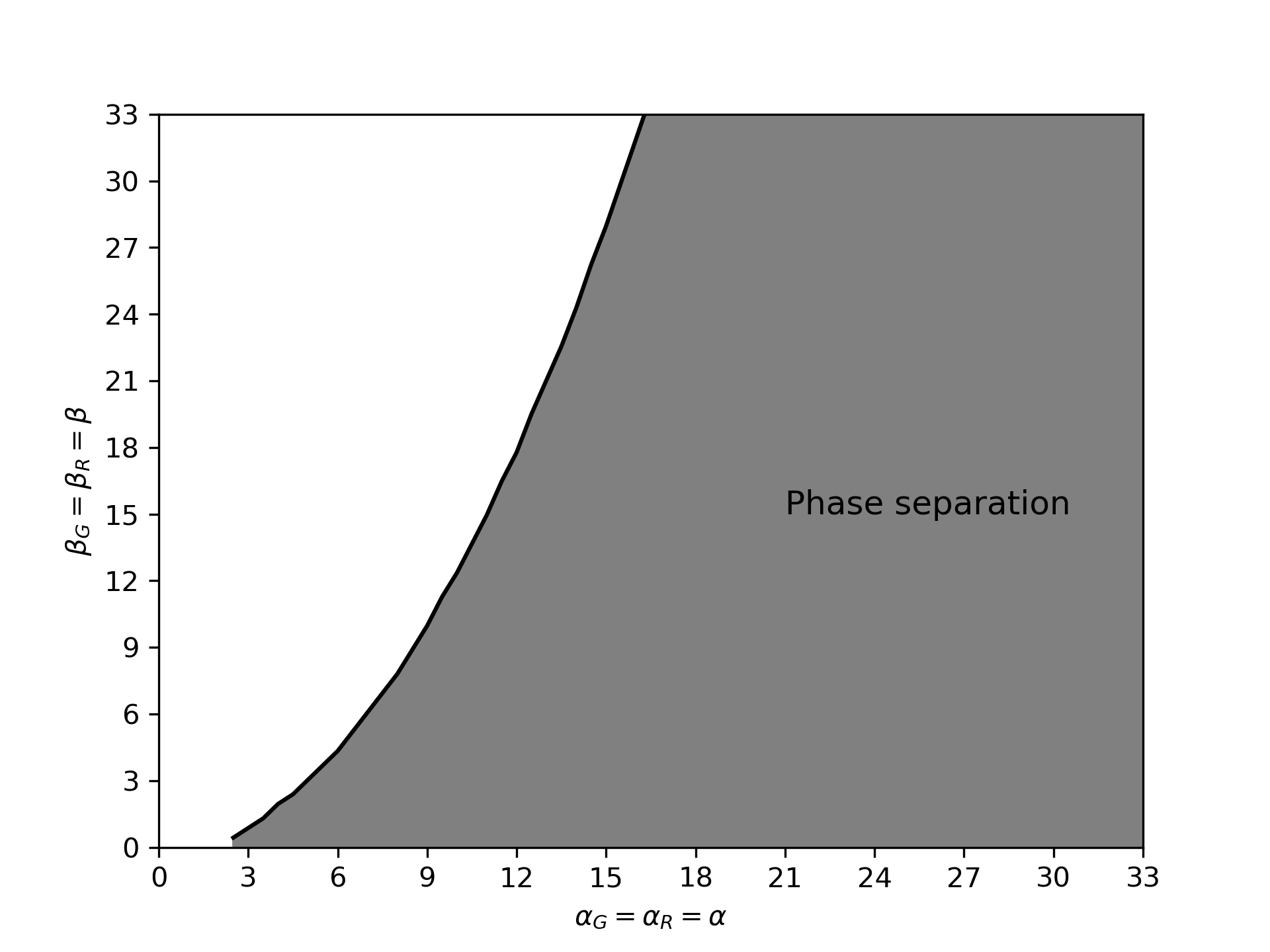}
\caption{Shaded region indicates $\alpha$-$\beta$  regime where phase separation occurs. $\alpha_G=\alpha_R=\alpha$ and $\beta_G=\beta_R=\beta$}.  
\label{fig:two_class_phase_separation_alpha_beta_regime}
\end{figure}  

The potential function is concave when the Hessian is negative semi-definite. The easiest way to check the negative semi-definiteness of the Hessian is to evaluate the sign of its eigenvalues. All the eigenvalues must be non-positive for a negative semi-definite matrix. Since we are interested in a non-concave potential function, our focus is on determining the parameter regime that leads to positive eigenvalues for the Hessian matrix.
Now, for any set of parameters $\alpha$, $\beta$ that result in a non-concave potential, there exists a region of densities where phase separation occurs. One of such density regions is shown for $\alpha_G=\alpha_R=\alpha=5$ and $\beta_G=\beta_R=\beta=0$ in Figure \ref{fig:two_class_phase_separation_alpha_5_beta_0_density_regime}. These densities were identified by evaluating the eigenvalues of the Hessian matrix. The points in the shaded region in Figure \ref{fig:two_class_phase_separation_alpha_5_beta_0_density_regime} represent densities of the two classes that result in at least one positive eigenvalue for the Hessian matrix.\\

Additionally, we performed a grid sweep in the $\alpha-\beta$ parameter regime, searching for a non-concave potential function. The points in the shaded region in Figure \ref{fig:two_class_phase_separation_alpha_beta_regime} represent ($\alpha, \beta$) values for which the potential function is non-concave in the density domain. The potential function is non-concave if the Hessian matrix has a positive eigenvalue for any pair of densities in the entire density domain.

\end{document}


\title{Supplementary Information}







\maketitle 
\section{Utility: two-class system}
Consider a class of agents identified as \textit{green agents} and another class of agents identified as \textit{red agents}. Their utilities are be defined as follows.

\begin{eqnarray}
    \tilde{h}_{G,i}(N_{G,i},N_{R,i})&=& \tilde{\alpha}_G N_{G,i}  - \tilde{\beta}_G N_{G,i}^2 - \ln N_{G,i} \nonumber\\
    & & + \ln(M - N_{G,i}-N_{R,i}) 
    \label{eq:utility_two-class_1}
\end{eqnarray}

\begin{eqnarray}
    \tilde{h}_{R,i}(N_{G,i},N_{R,i})&=& \tilde{\alpha}_R N_{R,i}  - \tilde{\beta}_R N_{R,i}^2 - \ln N_{R,i} \nonumber\\
    & & + \ln(M - N_{G,i}-N_{R,i}) 
    \label{eq:utility_two-class_2}
\end{eqnarray}
In Eq. \eqref{eq:utility_two-class_1}, and \eqref{eq:utility_two-class_2}, M is the number of cells in a block, $N_{G,i}$ and $N_{R,i}$ are respectively, the number of green agents and red agents in the $i^{th}$ block.
Eq. (\ref{eq:utility_two-class_1}) can be subjected to the following algebraic manipulations. 
\begin{eqnarray}
\tilde{h}_{G,i}(N_{G,i},N_{R,i})&=& \frac{\tilde{\alpha}_G N_{G,i} M}{M}  - \frac{\tilde{\beta}_G M^2 N_{G,i}^2}{M^2} - \ln N_{G,i} \nonumber\\
& & + \ln(M - N_{G,i}-N_{R,i}) \nonumber\\
\end{eqnarray}
Substituting for the densities, $\rho_{G,i} = \frac{N_{G,i}}{M}$ and $\rho_{R,i} = \frac{N_{R,i}}{M}$, 
\begin{eqnarray}
{h}_{G,i}(\rho_{G,i},\rho_{R,i})&=& \tilde{\alpha}_G M \rho_{G,i}   - \tilde{\beta}_G M^2 \rho_{G,i}^2 - \ln (M \rho_{G,i}) \nonumber\\
& & + \ln(M - M \rho_{G,i}- M \rho_{R,i}) \nonumber\\
\end{eqnarray}

\begin{eqnarray}
{h}_{G,i}(\rho_{G,i},\rho_{R,i})&=& \tilde{\alpha}_G M \rho_{G,i}   - \tilde{\beta}_G M^2 \rho_{G,i}^2 - \ln (\rho_{G,i}) \nonumber\\
& & -\ln M + \ln(M - M \rho_{G,i}- M \rho_{R,i}) \nonumber\\
\end{eqnarray}

\begin{eqnarray}
{h}_{G,i}(\rho_{G,i},\rho_{R,i})&=& \tilde{\alpha}_G M \rho_{G,i}   - \tilde{\beta}_G M^2 \rho_{G,i}^2 - \ln (\rho_{G,i}) \nonumber\\
& &  \ln(\frac{M - M \rho_{G,i}- M \rho_{R,i}}{M}) \nonumber\\
\end{eqnarray}

\begin{eqnarray}
{h}_{G,i}(\rho_{G,i},\rho_{R,i})&=& \alpha_G \rho_{G,i}   - \beta_G  \rho_{G,i}^2 - \ln \rho_{G,i} \nonumber\\
& & + \ln \left(1 -  \rho_{G,i}- \rho_{R,i} \right) \nonumber\\
\label{eqn_in_density}
\end{eqnarray}

where $\alpha_G = \tilde{\alpha}_G M$ and $\beta_G=\tilde{\beta}_G M^2$.\\

Similarly, 
\begin{eqnarray}
{h}_{R,i}(\rho_{G,i},\rho_{R,i})&=& \alpha_R \rho_{R,i}   - \beta_R  \rho_{R,i}^2 - \ln \rho_{R,i} \nonumber\\
& & + \ln \left(1 -  \rho_{G,i}- \rho_{R,i} \right) \nonumber\\
\label{eqn_in_density}
\end{eqnarray}

where $\alpha_R = \tilde{\alpha}_R M$ and $\beta_R=\tilde{\beta}_R M^2$.
\begin{widetext}
\section{Game-thoretic potential: two-class system}

The formulation lends itself to the use of the utility functional driven by density of agents in each cell-block $\rho_{G, i}, \rho_{R, i}$. 
\begin{eqnarray}
h_{R,i}(\rho_{G,i},\rho_{R,i})&=& \alpha_R \rho_{R,i}  - \beta_R \rho_{R,i}^2 - \ln \rho_{R,i} + \ln(1 - \rho_{G,i}-\rho_{R,i}) \\
    h_{G,i}(\rho_{G,i},\rho_{R,i})&=& \alpha_G \rho_{G,i}  - \beta_G \rho_{G,i}^2 - \ln \rho_{G,i}  + \ln(1 - \rho_{G,i}-\rho_{R,i}) 
\end{eqnarray}

We note that the potential $\phi({\rho_{G,i}, \rho_{R,i}})$ relates to this utility as below, 
\begin{eqnarray}
\frac{\partial \phi({\rho_{G,i}, \rho_{R,i}})}{\partial{\rho_{R,i}}} &=&  h_{R,i} (\rho_{G,i},\rho_{R,i}) 
    \label{eq:utility_definition_R}\\
    \frac{\partial \phi({\rho_{G,i}, \rho_{R,i}})}{\partial{\rho_{G,i}}} &=&  h_{G,i} (\rho_{G,i},\rho_{R,i}) \label{eq:utility_definition_G}
\end{eqnarray}

Integrating $h_{R,i}$ from Eq. \eqref{eq:utility_definition_R} with respect to $\rho_{R, i}$, we get a functional of the potential $\phi({\rho_{G,i}, \rho_{R,i}})$

\begin{eqnarray}
\phi &=& \sum_{i}^{Q} \int {h}_{R}({\rho_{G,i}, \rho_{R,i}}) d\rho_{R,i}  \nonumber\\ 
\int {h}_{R}({\rho_{G,i}, \rho_{R,i}}) d\rho_{R,i} &=&\frac{\alpha_R {\rho}^{2}_{R,i}}{2}   - \frac{\beta_R  \rho_{R,i}^3}{3} - {\rho_{R,i}} \ln \left(\rho_{R,i}\right)  + \rho_{R, i} - \left(1 - \rho_{G, i} - \rho_{R, i}\right)  \ln \left(1 -  \rho_{G,i} - \rho_{R,i}\right)\nonumber\\
& &+ (1 - \rho_{G, i} - \rho_{R, i}) + F(\rho_{G,i}) \nonumber\\
&=& \frac{\alpha_R {\rho}^{2}_{R,i}}{2}   - \frac{\beta_R  \rho_{R,i}^3}{3} - {\rho_{R,i}} \ln \left(\rho_{R,i}\right) - \left(1 - \rho_{G, i} - \rho_{R, i}\right) \ln \left(1 -  \rho_{G,i} - \rho_{R,i}\right) + (1 - \rho_{G, i}) \nonumber\\
& &+ F(\rho_{G,i}) \label{eq:potential_two_class}
\end{eqnarray}

where $F(\rho_{G, i})$ is a function independent of the variable $\rho_{R, i}$. Differentiating Eq. \eqref{eq:potential_two_class} with respect to $\rho_{G,i}$, \\
\begin{eqnarray}
\frac{\partial{\phi({\rho_{G,i}, \rho_{R,i}})}}{\partial{\rho_{G,i}}} & = & -\left(1  - \rho_{G, i} - \rho_{R, i}\right)\left(\frac{-1}{1 - \rho_{G,i} - \rho_{R, i}}\right) + \ln\left(1 - \rho_{G, i} - \rho_{R, i}\right) - 1 + F'(\rho_{G, i})\nonumber \\
&=& \ln\left(1 - \rho_{G, i} - \rho_{R, i}\right) + F'(\rho_{G, i})
\end{eqnarray}   

By definition (Eq. \eqref{eq:utility_definition_G}), 
\begin{eqnarray}
\frac{\partial{\phi({\rho_{G,i}, \rho_{R,i}})}}{\partial{\rho_{G,i}}} &=& h_{G, i}(\rho_{G, i}, \rho_{R,i})  \nonumber\\
\ln\left(1 - \rho_{G, i} - \rho_{R, i}\right) + F'(\rho_{G, i}) &=& \alpha_G \rho_{G,i}  - \beta_G \rho_{G,i}^2 - \ln \rho_{G,i} 
    + \ln(1 - \rho_{G,i}-\rho_{R,i})\nonumber\\
F'(\rho_{G, i}) &=& \alpha_G \rho_{G,i}  - \beta_G \rho_{G,i}^2 - \ln \rho_{G,i} \nonumber\\
F(\rho_{G, i}) &=& \frac{\alpha_G {\rho}^{2}_{G,i}}{2}   - \frac{\beta_G  \rho_{G,i}^3}{3} - {\rho_{G,i}} \ln \left(\rho_{G,i}\right)  + \rho_{G, i} + \text{Constant} \label{eq:F_rho_G}
\end{eqnarray}   

We thus get the potential functional in Eq.\eqref{eq:potential_two_class} by substituting $F(\rho_{G, i})$ from Eq. \eqref{eq:F_rho_G}, 

\begin{eqnarray}
\phi 
&=& \sum_{i}^{Q} \Bigg( \frac{\alpha_R {\rho}^{2}_{R,i}}{2}   - \frac{\beta_R  \rho_{R,i}^3}{3} - {\rho_{R,i}} \ln \left(\rho_{R,i}\right)  - \left(1 - \rho_{G, i} - \rho_{R, i}\right) \ln \left(1 - \rho_{G,i} - \rho_{R,i}\right) + (1 - \rho_{G, i}) \nonumber\\& &+ F(\rho_{G,i}) \nonumber\\
&=&\frac{\alpha_R {\rho}^{2}_{R,i}}{2}   - \frac{\beta_R  \rho_{R,i}^3}{3} - {\rho_{R,i}} \ln \left(\rho_{R,i}\right) 
 - \left(1 - \rho_{G, i} - \rho_{R, i}\right) \ln \left(1 -  \rho_{G,i} - \rho_{R,i}\right) + (1 - \rho_{G, i}) \nonumber \\
& &+  \frac{\alpha_G {\rho}^{2}_{G,i}}{2}   - \frac{\beta_G  \rho_{G,i}^3}{3} - {\rho_{G,i}} \ln \left(\rho_{G,i}\right)  + \rho_{G, i} + \text{Constant} \Bigg) \nonumber\\
\phi &=&\sum_{i}^{Q} \Bigg( \frac{\alpha_R {\rho}^{2}_{R,i}}{2}   - \frac{\beta_R  \rho_{R,i}^3}{3} +  \frac{\alpha_G {\rho}^{2}_{G,i}}{2}   - \frac{\beta_G  \rho_{G,i}^3}{3} - {\rho_{R,i}} \ln \left(\rho_{R,i}\right)  - {\rho_{G,i}} \ln \left(\rho_{G,i} \right) \nonumber\\
& &
- \left(1 - \rho_{G, i} - \rho_{R, i}\right) \ln \left(1 -  \rho_{G,i} - \rho_{R,i}\right) \Bigg)+ \text{Constant} 
\end{eqnarray}   
\end{widetext}

\section{Maximization of potential: Optimization problem}

In the previous section, we derived the potential from the utility formulation. Phase separation occurs when the potential is non-concave. In such cases, the system tries to maximize the potential by separating into multiple phases. In other words, the potential of the system in the phase separated configuration is more than that in the homogeneous configuration. \\

\subsection{One-class system}
For one-class system, our simulations showed that the system separates into two phases when the potential is non-concave. Here we compute the densities of the phases when the system separates into two phases in order to maximize its potential. To compute the densities, we first assume that the system separates into two phases, and then we formulate an optimization problem where the objective is to maximize the potential of the system, that includes the potential of both phases. If it turns out that at optimum the densities of two phases are equal, then that indicates system does not phase separate. On the other hand if the densities in the two phases are unequal, then that indicates the system phase separates. Let the overall density of the system be $\rho_0$ and the densities of the two phases formed after phase separation be $\rho_1$ and $\rho_2$. The potential of the system with a density $\rho$ is given by
\begin{eqnarray}
    \phi &=& \alpha \frac{\rho^2}{2} - \beta \frac{\rho^3}{3}- \rho \ln{\rho}- (1-\rho) \ln{(1-\rho)}
\end{eqnarray}
The total potential of the system when it phase separates into two phases with densities $\rho_1$ and $\rho_2$ is
\begin{eqnarray}
    \phi &=& n_1 \Bigg(\alpha \frac{\rho_{1}^2}{2} - \beta \frac{\rho_{1}^3}{3}- \rho_{1} \ln{\rho_{1}}- (1-\rho_{1}) \ln{(1-\rho_{1})}\Bigg) \nonumber \\
    & +& n_2\Bigg(\alpha \frac{\rho_{2}^2}{2} - \beta \frac{\rho_{2}^3}{3}- \rho_{2} \ln{\rho_{2}}- (1-\rho_{2}) \ln{(1-\rho_{2})}\Bigg) \nonumber \\
\end{eqnarray}
where $n_1$ and $n_2$ are the volume fraction of each phases. Note that $n_1+n_2=1$.

The potential maximization problem can be formally represented as follows.

\begin{eqnarray}
   \max \ \phi &=& n_1 \Bigg(\alpha \frac{\rho_{1}^2}{2} - \beta \frac{\rho_{1}^3}{3}- \rho_{1} \ln{\rho_{1}}- (1-\rho_{1}) \ln{(1-\rho_{1})}\Bigg) \nonumber \\
    & +& n_2\Bigg(\alpha \frac{\rho_{2}^2}{2} - \beta \frac{\rho_{2}^3}{3}- \rho_{2} \ln{\rho_{2}}- (1-\rho_{2}) \ln{(1-\rho_{2})}\Bigg) \nonumber \\
\end{eqnarray}
\begin{eqnarray}
    \text{Subject  to}: &  & \nonumber \\
    n_1 \rho_1 + n_2 \rho_2 &=&\rho_0 \\
    n_1+n_2&=&1
\end{eqnarray}

We implemented the potential maximization problem in Python and solved using the \textit{ipopt} method of the package \textit{pyomo}. The optimization was performed at the same parameter settings that were used in the agent based simulations (displayed in Figure 5 in the main manuscript). The results are provided in Table \ref{Potential_maximization_results_one_class}.  Reader may recall that no phase separation was observed in the simulations for $\rho_0=0.1$ for any $\alpha$ (Refer to Figure 5 A,B,C). The equilibrium density was equal to the overall density. The potential maximization problem provides identical results. Once can note in Table \ref{Potential_maximization_results_one_class} (No. 1,2,3) that for $\rho_0=0.1$, and $\alpha=0,4,8$, both the densities $\rho_1$ and $\rho_2$ that maximized the potential are equal to 0.1. This means that there is no phase separation.  Since both phases have same density, the volume fractions of the two phases are irrelevant and any volume fractions $n_1$ and $n_2$ would result in the same potential. In summary, the optimization results are in agreement with the agent-based simulations. This is not surprising because, the potential is concave at these points and therefore phase separation is not preferred. Similarly, for $\rho_0=0.25$ and $0.5$, phase separation was not observed in the simulations as well as in the optimization when $\alpha=0$, and $4$ (Refer to configurations D,E,G , and H in Figure 5).  However, for $\rho_0=0.25, 0.5$ and $\alpha=8$, phase separation was observed in the simulations (Refer to configurations F and I in Figure 5). For $\rho_0=0.25$ and $\alpha=8$, densities of the two phases observed in the simulations are $\rho^{*}_1=0.039$ and $\rho^{*}_2=0.821$. Our optimization problem predicts phase separation for this parameter set. However, there are minor differences in predicted densities of the two phases - $\rho_1=0.021$ and $\rho_2=0.979$. One possible reason for the mismatch could be the integer requirements for the number of blocks of each phase in the simulations. In simulations, the number of blocks in the two phases are respectively $28$ and $8$. If we compute the number of blocks from the fraction of each phases obtained by the optimizer, that would be $0.761*36=27.396$ and $0.239*36=8.604$. The mismatch in the observed and predicted densities could be due to the automatic adjustment of the densities in the simulations to have integer number blocks for each phase. Similar analysis can be performed for the case of $\rho_0=0.5$ and $\alpha=8$.

\begin{table*}[]
    \centering
    \caption{Solution to the potential maximization problem: Densities in two phases and the volume fraction of each phase}
\begin{tabular}{|p{1cm}|p{2cm}|p{1cm}|p{1cm}|p{1cm}|p{1cm}|p{1cm}|p{1cm}|p{1cm}|p{1cm}|p{1cm}|p{1cm}|p{1cm}|}
        \hline
        &    &  &    &  &  \multicolumn{4}{l|}{Optimization Results} & \multicolumn{4}{l|}{Simulation Results}\\ [1em]\cline{6-13} 
          No. &  Configuration  & $\alpha$ &  $\beta$  &$\rho_{0}$  &  $\rho_1$ &  $\rho_2$ & $n_{1}$ & $n_{2}$ &$\rho^{*}_1$ &  $\rho^{*}_2$ & $n_{1}$ & $n_{2}$ \\ 
         
         & Figure 5 &  &  & &  & & & & & & &\\ 
         \hline
         1 & A &0 & 0 & 0.1 & 0.1& 0.1 &0.527 & 0.473 &0.1 &0.1 & -& -\\[0.5em]
         2 & B&4 & 0 & 0.1 & 0.1& 0.1 & 0.505& 0.495 &0.1 & 0.1& -&- \\[0.5em]
         3 & C & 8 & 0 & 0.1 &  0.1& 0.1&0.488 & 0.512 & 0.1&0.1 & -&-\\[0.5em]
         4 & D & 0 & 0 & 0.25 & 0.25& 0.25 & 0.506&0.494 & 0.25& 0.25& -&-\\[0.5em]
         5 & E & 4 & 0 & 0.25 &0.25 & 0.25 & 0.492&0.508 & 0.25& 0.25 & -&-\\[0.5em]
         6 & F& 8 & 0 & 0.25 &0.021  & 0.979&0.761& 0.239 &  0.039& 0.988& 0.778&0.222\\[0.5em]
        7 & G & 0 & 0 & 0.5 &0.5 &0.5  & 0.681& 0.319 &0.5 &0.5 & -&-\\ [0.5em]
         8 & H &4 & 0 & 0.5 & 0.5& 0.5 & 0.5& 0.5 & 0.5& 0.5& -&-\\[0.5em]
         9 & I &8 & 0 & 0.5 & 0.021 & 0.979& 0.5& 0.5 & 0.061& 0.991&0.528 & 0.472\\[0.5em]
         \hline
        
    \end{tabular}
    \label{Potential_maximization_results_one_class}
\end{table*}
Densities of the two phases that maximizes the potential are called \textit{binodal densities}. For $\alpha=8$ and $\beta=0$, the densities $\rho_1=0.021$ and  $\rho_2=0.979$ are the binodal densities reported in Figure 2 in the manuscript. 

\subsection{Two-class system}

In our simulations, we observed that two class system separates into three phases. Consider the separation of a uniform mixture formed of two classes of agents into three phases. Through formulating and solving a potential maximization problem, we show that the system phase separates in order to increase its potential. Let the overall densities of the two classes be $\rho_{G0}$ and $\rho_{R0}$. The densities of the two classes in the three phases after phase separation are denoted by $\rho_{G,I}$,$\rho_{R,I}$,$\rho_{G,II}$,$\rho_{R,II}$, $\rho_{G,III}$ and $\rho_{R,III}$. Volume fractions of the phases are represented by $n_I$,$n_{II}$ and $n_{III}$. 
\begin{widetext}

\begin{eqnarray}
    Max \ & \phi = n_I \Bigg( \frac{\alpha_R {\rho}^{2}_{R,I}}{2}   - \frac{\beta_R  \rho_{R,I}^3}{3} +  \frac{\alpha_G {\rho}^{2}_{G,I}}{2}   - \frac{\beta_G  \rho_{G,I}^3}{3} 
   - {\rho_{R}} \ln \left(\rho_{R,I}\right)  - {\rho_{G,I}} \ln \left(\rho_{G,I} \right)  
- \left(1 - \rho_{G, I} - \rho_{R, I}\right) \ln \left(1 -  \rho_{G,I} - \rho_{R,I}\right) \Bigg) \nonumber \\
&+ n_{II}\Bigg( \frac{\alpha_R {\rho}^{2}_{R,II}}{2}   - \frac{\beta_R  \rho_{R,II}^3}{3} +  \frac{\alpha_G {\rho}^{2}_{G,II}}{2}   - \frac{\beta_G  \rho_{G,II}^3}{3}  - {\rho_{R}} \ln \left(\rho_{R,II}\right)  - {\rho_{G,II}} \ln \left(\rho_{G,II} \right)\nonumber \\
&- \left(1 - \rho_{G, II} - \rho_{R, II}\right) \ln \left(1 -  \rho_{G,II} - \rho_{R,II}\right) \Bigg) \nonumber \\
&+ n_{III}\Bigg( \frac{\alpha_R {\rho}^{2}_{R,III}}{2}   - \frac{\beta_R  \rho_{R,III}^3}{3} +  \frac{\alpha_G {\rho}^{2}_{G,III}}{2}   - \frac{\beta_G  \rho_{G,III}^3}{3}  - {\rho_{R}} \ln \left(\rho_{R,III}\right)  - {\rho_{G,III}} \ln \left(\rho_{G,III} \right)\nonumber \\
&- \left(1 - \rho_{G, III} - \rho_{R, III}\right) \ln \left(1 -  \rho_{G,III} - \rho_{R,III}\right) \Bigg)
\label{objective}
\end{eqnarray}

\begin{eqnarray}
\text{Subject  to} :  &  & \nonumber \\  \nonumber\\
 \left(n_I \rho_{G,I} + n_{II}  \rho_{G,II}+ n_{III}   \rho_{G,II} \right) 
 &=& \rho_{G0} 
 \label{green_conservation}
\end{eqnarray}
 \begin{eqnarray}
     (n_I  \rho_{R,I} + n_{II}  \rho_{R,II}+ n_{III}   \rho_{R,II}) 
   & = &\rho_{R0}
   \label{red_conservation}
 \end{eqnarray}
    \begin{eqnarray}
        n_I +n_{II} +n_{III}=1 
        \label{volume_conservation}
    \end{eqnarray}
\end{widetext}

Eq. \eqref{objective} provides potential of the system which includes the potential of all the three phases. This potential is the objective that needs to be maximized. While maximizing the potential, the amount or the number of agents should be conserved. Eq. \eqref{green_conservation} and \eqref{red_conservation} together with Eq. \eqref{volume_conservation} conserves the number of agents of each class.
If there is no phase separation, densities that maximizes the potential should be same as the overall density of each class, i.e., $\rho_{G,I}=\rho_{G,II}=\rho_{G,III}=\rho_{G0}$ and $\rho_{R,I}$=$\rho_{R,II}=\rho_{R,III}=\rho_{R0}$. When phase separation occurs, the densities of the two classes in the three phases will be unequal. 
The above optimization problem was solved using the method \textit{ipopt} from the Python package, \textit{pyomo}. 

The optimization problem is solved for parameters provided in Table II in the manuscript. Solution to the optimization problem is provided in Table \ref{Potential_maximization_results}.  For $\alpha=0$, and $2.023$, phase separation was not observed in the simulations. The potential maximization problem also predicts \textit{no phase separation}. This is evident from the fact that the densities of the two classes are equal to the overall densities in each phase. Similarly, as observed in the simulations, the potential maximization problem also predicts \textit{no phase separation} for $\alpha=5$ at $\rho_{G0}=\rho_{R0}=0.1$. For $\alpha=5$ and $\rho_{G0}=\rho_{R0}=0.25$, solution to the potential maximization problem indicates phase separation. The predicted densities in the three phases are $\rho_{G,I}=0.863$, $\rho_{R,I}=0.012$,   $\rho_{G,II}=0.012$, $\rho_{R,II}=0.863$, $\rho_{G,III}=0.131$ and $\rho_{R,III}=0.131$. These exactly match with the results observed in the simulations. The fraction of each phase predicted are 0.194, 0.194 and 0.613. Then, the number of blocks predicted from these fractions are $0.194 \times 36 =6.984 \approx 7$, $0.194 \times 36 =6.984 \approx 7$ and $0.613 \times 36 =22.067 \approx 22$. The number of blocks observed in the simulations are respectively, $7,7$, and $22$.  In summary, the solution to the optimization problem exactly match with the observations from the simulations. \\

For $\rho_{G0}=\rho_{R0}=0.4$, solution to the potential maximization problem indicates phase separation. The predicted densities in the three phases are $\rho_{G,I}=0.863$, $\rho_{R,I}=0.012$,   $\rho_{G,II}=0.012$, $\rho_{R,II}=0.863$, $\rho_{G,III}=0.131$ and $\rho_{R,III}=0.131$. But these predictions are different from the equilibrium density values observed in the simulations. The equilibrium densities observed in the simulation are, $\rho^{*}_{G,I}=0.887$, $\rho^{*}_{R,I}=0.011$,   $\rho^{*}_{G,II}=0.011$, $\rho^{*}_{R,II}=0.887$, $\rho^{*}_{G,III}=0.154$ and $\rho^{*}_{R,III}=0.154$.  One possible reason for the mismatch could be the integer requirements for the number of blocks of each phase. The fraction of each phase predicted are $0.439, 0.439$ and $0.122$. The number of blocks predicted from these fractions are $0.439 \times 36 =15.804$, $0.439 \times 36 =15.804$ and $0.122 \times 36 =4.392$. The number of blocks observed in the simulations are $15,15$, and $6$,    respectively.  In the simulations, the requirement that the number of blocks of each phase is an integer is a hard constraint.  To ensure this, the densities of each phase adjust automatically. Therefore the densities can be different from the theoretical prediction. 
\begin{table*}[!ht]
    \centering
    \caption{Solution to the potential maximization problem: Densities of the two classes in three phases and the volume fraction of each phase}
    \begin{tabular}{|c|c|c|c|c|c|c|c|c|c|c|c|c|c|}
        \hline

        \hline
         \multirow{2}{*} {No.} &  \multirow{2}{*}{$\alpha_G=\alpha_R=\alpha$ }& \multirow{2}{*}{$\beta_G=\beta_R=\beta$ } &\multirow{2}{*}{$\rho_{G0}$} &\multirow{2}{*}{$\rho_{R0}$} &  \multicolumn{2}{c|}{{Phase-I}} & \multicolumn{2}{c|}{{Phase-II}}  & \multicolumn{2}{c|}{{Phase-III}} & \multirow{2}{*}{$n_{I}$} & \multirow{2}{*}{$n_{II}$ }&\multirow{2}{*}{$n_{III}$}\\  \cline{6-11} 
          &  & & & &$\rho^{*}_{G,I}$ & $\rho^{*}_{R,I}$ & $\rho^{*}_{G,II}$& $\rho^{*}_{R,II}$ & $\rho^{*}_{G,III}$& $\rho^{*}_{R,III}$ &  &  &  \\ 
         \\ [-1em]
         \hline
         1& 0 & 0 & 0.1 & 0.1 & 0.1 &0.1 & 0.1 & 0.1 & 0.1 & 0.1 & 0.267 &0.263 &0.469 \\[0.5em]
         2& 2.023 & 0 & 0.1 & 0.1 & 0.1 &0.1 & 0.1 & 0.1 & 0.1 & 0.1 & 0.272& 0.278 &0.450 \\[0.5em]
         3& 5 & 0 & 0.1 & 0.1 & 0.1 &0.1 & 0.1 & 0.1 & 0.1 & 0.1 & 0.363 & 0.313 & 0.324 \\[0.5em]
           
           4& 0 & 0 & 0.25 & 0.25 & 0.25 & 0.25 & 0.25 & 0.25 & 0.25 & 0.25 & 0.276& 0.276 & 0.447\\[0.5em]
           5& 2.023 & 0 & 0.25 & 0.25 & 0.25 & 0.25 & 0.25 & 0.25 & 0.25 & 0.25 & 0.283 & 0.376 & 0.341\\[0.5em]
           6& 5 & 0 & 0.25 & 0.25 & 0.863 &0.012&   0.012 & 0.863& 0.131 & 0.131 &  0.194 &0.194  &0.613 \\[0.5em]
           
           7& 0 & 0 & 0.4 & 0.4 & 0.4 & 0.4 & 0.4 & 0.4 & 0.4 & 0.4  & 0.276& 0.276 & 0.447\\[0.5em]
           8& 2.023 & 0 & 0.4 & 0.4 & 0.4& 0.4 & 0.4 & 0.4 & 0.4 & 0.4 & 0.276& 0.276 & 0.448\\[0.5em]
           9& 5 & 0 & 0.4 & 0.4 & 0.863 &0.012&   0.012 & 0.863& 0.131 & 0.131 &  0.439 & 0.439 &0.122 \\[0.5em]
        \hline
        
    \end{tabular}
    \label{Potential_maximization_results}
\end{table*}

\section{Solution to cubic equation \label{cubic_solution}}
A general cubic equation can be represented as the following.
\begin{eqnarray}
    a x^3+bx^2+cx+d=0
\end{eqnarray}
Three parameters, $\Delta_0$, $\Delta_1$, and $C$ are defined as  follows
\begin{eqnarray}
    \Delta_0&=&b^2-3ac \\
    \Delta_1 &=& 2b^3-9abc+27a^2d\\
    C&=& \sqrt[3]{\frac{\Delta_1 \pm \sqrt{{\Delta_1}^2-4{\Delta_0}^3}}{2}}
\end{eqnarray}

The solutions to the cubic equation can be derived from Cardano's formula and these solutions are provided in Eq. \eqref{cubic_solution}.
\begin{eqnarray}
    x_k&=& -\frac{1}{3a}\left(b+\xi^kC+\frac{\Delta_0}{\xi^kC}\right), k \in \{ 0,1,2 \}
    \label{cubic_solution}
\end{eqnarray}
where
\begin{eqnarray}
    \xi&=& \frac{-1+\sqrt{-3}}{2}
\end{eqnarray}

\subsection*{Transition density region}
\begin{figure*}[]
\centering 
\includegraphics[width=\linewidth]{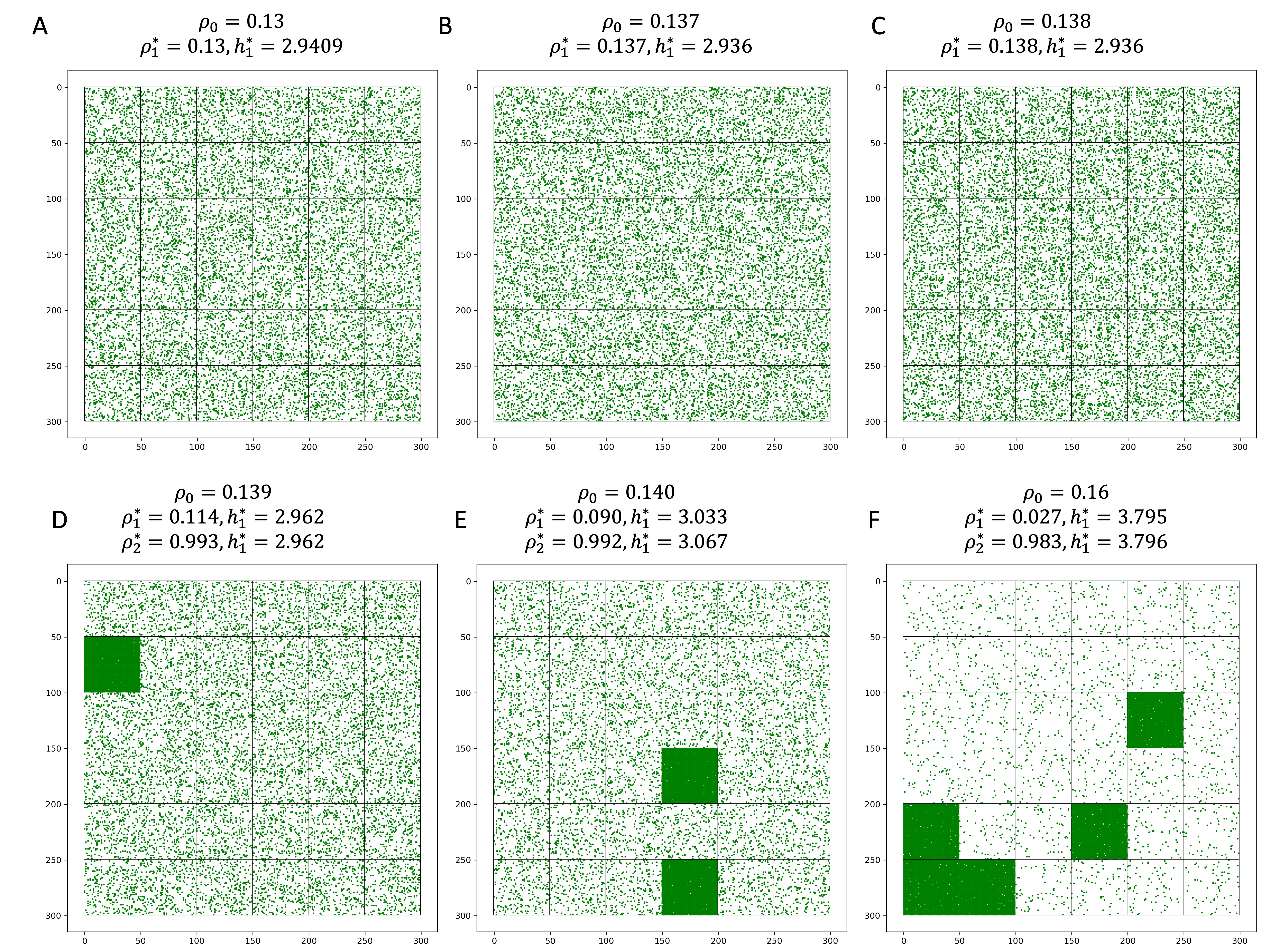}
\caption{$\alpha=8, \beta=0$. Agent configurations in the transition density region}
\label{fig:one_class_critical_density}
\end{figure*}

We have already described that the phase separation occurs when the density lies between the spinodal densities. Let us look at a phase transition region through agent-based simulations. Agent configurations for six densities are provided in Figure \ref{fig:one_class_critical_density}. The parameters used in the simulations are $\alpha=8$ and $\beta=0$. The lower and upper spinodal densities corresponding to these parameters are 0.146 and 0.854, respectively. The overall densities of each experiment is represented by the variable $\rho_0$. For instance, Figure \ref{fig:one_class_critical_density}A represents the equilibrium configuration of agents in an experiment with  $\rho_0=0.130$, indicating that there are $11,700$ agents $(0.130 \times 90,000)$. Initially, we populate these agents in the lattice, one by one, by placing each agent in  a vacant cell randomly sampled from the entire lattice. This process distributes the agents uniformly in the lattice. However, because of random assignment, one cannot expect each block to be populated with an exact density of $\rho_0$. However, the densities of all the blocks will be around $\rho_0$, meaning, some blocks with density less than $\rho_0$ and remaining with more than $\rho_0$. Besides, for a given $\rho_0$, the density distribution slightly changes in each realisation of the experiment because the agents are randomly populated. \\

When all the cells have an initial density less that the lower spinodal density ($0.146$), agents do not phase-separate. This behavior is noticed for $\rho_0=0.130$ as shown in Figure \ref{fig:one_class_critical_density}A. The initial densities and corresponding utilities in each block are provided in Table \ref{tab:InitialConfig_A}. It can be noted that all blocks have density less than the lower spinodal density. \\

As we increase $\rho_0$, a few blocks are populated with a density higher than the lower spinodal density and the remaining blocks with a density lower than that.  The initial density in each block for $\rho_0=0.137$ and $\rho_0=0.139$ are provided in Table \ref{tab:InitialConfig_B} and \ref{tab:InitialConfig_D}. There are a few blocks with density more than $0.146$ in each of these cases.  In such cases, the final equilibrium configuration depends on the sequence of movements made by the agents. Figure \ref{fig:one_class_critical_density}B shows a realisation where the population does not phase-separate. However, we have observed phase-separation in other realisation of the experiment with the same $\rho_0=0.137$. Similarly, Figure \ref{fig:one_class_critical_density}C shows equilibrium configuration for $\rho_0=0.138$. Again, this is only one realisation of the experiment, we have observed phase separation in other realisations  for $\rho_0=0.138$.  In  Figure \ref{fig:one_class_critical_density}D and E, we show the equilibrium configuration for $\rho_0=0.139$ and $\rho_0=0.140$. In these experiments the population is observed to phase-separate. The trend we observe is, as $\rho_0$ increases, the number of blocks with a density higher than lower spinodal density increases. With more number of blocks with an initial density larger than the lower spinodal density, the chances of phase separation also increases. \\

Finally, initial configurations where every block has a density between the spinodal densities phase-separate without exception. Equilibrium configuration in one such experiment is shown in Figure \ref{fig:one_class_critical_density}F. The corresponding initial densities are provided in Table \ref{tab:InitialConfig_F}. Note, all the blocks have initial densities within the spinodal densities. In such cases, irrespective of the sequence of movement made by the agents, the population will phase-separate. 

\begin{table}[!ht]
\caption{Initial configuration of agents for $\rho_0=0.130$}
\centering
    \begin{tabular}{|c|c|c|c|c|c|} 
        \hline
         Block & Density & Utility & Block & Density & Utility \\
        \hline
        1 & 0.134 & 2.9380 &19 & 0.138 & 2.9360\\
        2 & 0.1264 & 2.9444&20 & 0.126 & 2.9448 \\
        3 & 0.1392 & 2.9356 &21 & 0.1324 & 2.9391\\
        4 & 0.1336 & 2.9383 &22 & 0.1284 & 2.9424\\
        5 & 0.1312 & 2.9400 &23 & 0.1396 & 2.9354\\
        6 & 0.1288 & 2.9420 &24 & 0.1288 & 2.9420\\
        7 & 0.1372 & 2.9363 &25 & 0.1344 & 2.9378 \\
        8 & 0.1272 & 2.9435 &26 & 0.1228 & 2.9486\\
        9 & 0.1252 & 2.9457 &27 & 0.1332 & 2.9386\\
        10 & 0.1328 & 2.9388 &28 & 0.136 & 2.9370\\
        11 & 0.1232 & 2.9481 &29 & 0.128 & 2.9428\\
        12 & 0.1196 & 2.9530 &30 & 0.1284 & 2.9424\\
        13 & 0.1324 & 2.9391 &31 & 0.12 & 2.9524 \\
        14 & 0.1396 & 2.9354 &32 & 0.1348 & 2.9376 \\
        15 & 0.1356 & 2.9371 &33 & 0.1312 & 2.9400 \\
        16 & 0.1204 & 2.9518 &34 & 0.1128 & 2.9650\\
        17 & 0.1292 & 2.9417 &35 & 0.1356 & 2.9371\\
        18 & 0.1212 & 2.9507 &36 & 0.1328 & 2.9388\\
        \hline
    \end{tabular}

\label{tab:InitialConfig_A}
\end{table}

   \begin{table}[!ht]
   \caption{Initial configuration of agents for $\rho_0=0.137$}
    \centering
    \begin{tabular}{|c|c|c|c|c|c|} 
        \hline
        Block & Density & Utility & Block & Density & Utility \\
        \hline
        1 & 0.1292 & 2.9416 & 19 & 0.1268 & 2.9440 \\
        2 & 0.1492 & 2.9345 & 20 & 0.1428 & 2.9346 \\
        3 & 0.1444 & 2.9344 & 21 & 0.1432 & 2.9346 \\
        4 & 0.12 & 2.9524 & 22 & 0.1288 & 2.9420 \\
        5 & 0.136 & 2.9369 & 23 & 0.14 & 2.9353 \\
        6 & 0.144 & 2.9345 & 24 & 0.1436 & 2.9345 \\
        7 & 0.13 & 2.9410 & 25 & 0.1316 & 2.9397 \\
        8 & 0.1488 & 2.9344 & 26 & 0.1324 & 2.9391 \\
        9 & 0.1348 & 2.9376 & 27 & 0.13 & 2.9410 \\
        10 & 0.1396 & 2.9354 & 28 & 0.1388 & 2.9357 \\
        11 & 0.1336 & 2.9383 & 29 & 0.1368 & 2.9365 \\
        12 & 0.1372 & 2.9363 & 30 & 0.1364 & 2.9367 \\
        13 & 0.138 & 2.9360 & 31 & 0.1224 & 2.9491 \\
        14 & 0.1324 & 2.9391 & 32 & 0.142 & 2.9348 \\
        15 & 0.1348 & 2.9376 & 33 & 0.1404 & 2.9352 \\
        16 & 0.1404 & 2.9352 & 34 & 0.1296 & 2.9413 \\
        17 & 0.1356 & 2.9371 & 35 & 0.1508 & 2.9347 \\
        18 & 0.14 & 2.9353 & 36 & 0.1476 & 2.9343 \\
        \hline
    \end{tabular}
    
    \label{tab:InitialConfig_B}
\end{table}

\begin{table}[!ht]
\caption{Initial configuration of agents for $\rho_0=0.139$}
    \centering
    \begin{tabular}{|c|c|c|c|c|c|} 
        \hline
        Block & Density & Utility & Block & Density & Utility \\
        \hline
        1 & 0.134 & 2.9380 & 19 & 0.1384 & 2.9358 \\
        2 & 0.1388 & 2.9357 & 20 & 0.1344 & 2.9378 \\
        3 & 0.1456 & 2.9343 & 21 & 0.15 & 2.9346 \\
        4 & 0.1424 & 2.9347 & 22 & 0.15 & 2.9346 \\
        5 & 0.1332 & 2.9386 & 23 & 0.1308 & 2.9403 \\
        6 & 0.134 & 2.9380 & 24 & 0.1272 & 2.9435 \\
        7 & 0.154 & 2.9356 & 25 & 0.1432 & 2.9346 \\
        8 & 0.1304 & 2.9406 & 26 & 0.1404 & 2.9352 \\
        9 & 0.1388 & 2.9357 & 27 & 0.1412 & 2.9349 \\
        10 & 0.1412 & 2.9350 & 28 & 0.1448 & 2.9344 \\
        11 & 0.126 & 2.9448 & 29 & 0.14 & 2.9353 \\
        12 & 0.136 & 2.9369 & 30 & 0.1308 & 2.9403 \\
        13 & 0.142 & 2.9347 & 31 & 0.1464 & 2.9343 \\
        14 & 0.12 & 2.9524 & 32 & 0.1432 & 2.9346 \\
        15 & 0.1424 & 2.9347 & 33 & 0.1328 & 2.9388 \\
        16 & 0.144 & 2.9345 & 34 & 0.1436 & 2.9345 \\
        17 & 0.1384 & 2.9358 & 35 & 0.1456 & 2.9343 \\
        18 & 0.1464 & 2.9343 & 36 & 0.1336 & 2.9383 \\
        \hline
    \end{tabular}
    
    \label{tab:InitialConfig_D}
\end{table}

\begin{table}[!ht]
\caption{Initial configuration of agents for $\rho_0=0.160$}
\centering
\begin{tabular}{|c|c|c|c|c|c|}
\hline
 Block& Density & Utility & Block& Density & Utility \\
\hline
        1 & 0.1524 & 2.9351 & 19 & 0.1648 & 2.9413 \\
        2 & 0.1636 & 2.9405 & 20 & 0.1608 & 2.9387\\
        3 & 0.1628 & 2.9399 &21 & 0.1600 & 2.9382 \\
        4 & 0.1584 & 2.9374 &22 & 0.1624 & 2.9397\\
        5 & 0.1508 & 2.9347 &23 & 0.1604 & 2.9385\\
        6 & 0.1668 & 2.9429 &24 & 0.1664 & 2.9426\\
        7 & 0.1552 & 2.9360 &25 & 0.1708 & 2.9464\\
        8 & 0.1508 & 2.9347 &26 & 0.1608 & 2.9387\\
        9 & 0.1540 & 2.9356 &27 & 0.1524 & 2.9351\\
        10 & 0.1544 & 2.9357 & 28 & 0.1704 & 2.9460\\
        11 & 0.1552 & 2.9360 &29 & 0.1492 & 2.9345\\
        12 & 0.1536 & 2.9354 &30 & 0.1608 & 2.9387\\
        13 & 0.1552 & 2.9360 &31 & 0.1756 & 2.9512\\
        14 & 0.1612 & 2.9389 &32 & 0.1648 & 2.9413\\
        15 & 0.1560 & 2.9363 &33 & 0.1656 & 2.9419\\
        16 & 0.1592 & 2.9378 &34 & 0.1572 & 2.9368\\
        17 & 0.1740 & 2.9495 &35 & 0.1640 & 2.9408\\
        18 & 0.1612 & 2.9389 &36 & 0.1488 & 2.9344\\
\hline
\end{tabular}
\label{tab:InitialConfig_F}
\end{table}